\documentclass[12pt,preprint]{aastex}

\usepackage{bm}
\newcommand{\symvec}[1]{\mbox{\boldmath${#1}$}}

\newcommand{\bvec}[1]{\textbf{#1}}

\shorttitle{PCA OF ACS/WFC PSFs}
\shortauthors{Jee et al.}

\begin{document}

\title{PRINCIPAL COMPONENT ANALYSIS OF THE TIME- AND POSITION-DEPENDENT POINT SPREAD FUNCTION OF 
THE ADVANCED CAMERA FOR SURVEYS} 

\author{M.J. JEE\altaffilmark{1,4},
        J.P. BLAKESLEE\altaffilmark{2},
	M. SIRIANNI\altaffilmark{3},
        A.R. MARTEL\altaffilmark{1},
	R.L. WHITE\altaffilmark{3},
        AND
	H.C. FORD\altaffilmark{1}
}

\begin{abstract}
We describe the time- and position-dependent point spread function (PSF) variation of the Wide
Field Channel (WFC) of the Advanced Camera for Surveys (ACS) with the principal component analysis (PCA) 
technique. 
The time-dependent change is caused by the temporal variation of the $HST$ focus whereas
the position-dependent PSF variation in ACS/WFC at a given focus
is mainly the result of changes 
in aberrations and charge diffusion across the detector, which
appear as position-dependent changes in elongation of the astigmatic core and blurring of the PSF, respectively.
Using $>400$ archival images of star cluster fields, we construct a ACS PSF library
covering diverse environments of the $HST$ observations (e.g., focus values).
We find that interpolation of a small number ($\sim20$) of 
principal components or ``eigen-PSFs'' per exposure can robustly reproduce the 
observed variation of the ellipticity and size of the PSF. Our primary interest in this investigation
is the application of this PSF library to precision weak-lensing analyses, where accurate knowledge of the 
instrument's PSF is crucial. However, the high-fidelity of the model judged from 
the nice agreement with observed PSFs
suggests that the model is potentially also useful in other applications such as
crowded field stellar photometry, galaxy profile fitting, AGN studies, etc., which similarly
demand a fair knowledge of the PSFs at objects' locations.
Our PSF models, applicable to any WFC image rectified with the Lanczos3 kernel,
are publicly available.
\end{abstract}

\keywords{Astronomical Instrumentation ---
Astronomical Techniques ---
Data Analysis and Techniques ---
Astrophysical Data ---
Star Clusters and Associations}

\altaffiltext{1}{Department of Physics and Astronomy, Johns Hopkins
University, 3400 North Charles Street, Baltimore, MD 21218.}
\altaffiltext{2}{Department of Physics and Astronomy, Washington State University, Pullman, WA 99164}
\altaffiltext{3}{STScI, 3700 San Martin Drive, Baltimore, MD 21218.}
\altaffiltext{4}{Department of Physics, University of California, One Shields Avenue, Davis, CA 95616}

\section{INTRODUCTION \label{section_introduction}}
Even in the absence of atmospheric turbulence, the finite aperture of Hubble Space Telescope ($HST$) causes
light from a point source to spread at the focal plane with the diffraction pattern mainly
reflecting the telescope's aperture and optical path difference function. 
Although the point-spread-function (PSF)
of $HST$ is already far smaller than what one can achieve with any of the current ground-based facilities, 
astronomers' endless efforts to push to the limits of their scientific observations with $HST$
ever increase the demand for the better knowledge of the instrument's PSF.
Especially, since the installation of the Advanced Camera for Surveys (ACS) on $HST$, there
have been concentrated efforts to carefully monitor and understand the instrument's PSFs, and to
utilize the unparalleled resolution and sensitivity of ACS in gravitational weak-lensing 
(e.g., Jee et al. 2005a; Heymans et al. 2005; Schrabback et al. 2007; Rhodes et al. 2007).

Modeling the PSFs of ACS has proven to be non-trivial because of its complicated time- and position-dependent
variation. The time-dependent change occurs due to the variation in the $HST$ focus, which
relates to the constant shrinking of the secondary mirror truss structure and the thermal 
breathing of $HST$. The former is the main cause of the long-term focus change, and the secondary mirror position has been
occasionally adjusted to compensate for this shrinkage (Hershey 1997). The latter is responsible for
the short-term variation of the $HST$ focus and is affected by the instrument's earth heating, sun angle, prior pointing
history, roll angle, etc. Even at a fixed focus value of $HST$, the PSFs of ACS also significantly change across the 
detector from the variation of the CCD thickness and the focal plane errors, which
appear as position-dependent changes in charge diffusion and elongation of the astigmatic cores, respectively.

The strategies to model these PSF variations can be categorized into two types: an empirical
approach based on real stellar field observations and a theoretical prediction based on
the understanding of the instrument's optics. The first method treats the optical system
of the instrument nearly as a blackbox and mainly draws information from observed
stellar images. Although the PSF variation pattern can be most straightforwardly described 
by the variation of the
pixel intensity as a function of position (e.g., Anderson \& King 2006), frequently orthogonal expansion of the observed
PSFs (e.g., Lauer 2002; Bernstein \& Jarvis 2002; Refregier 2003) have been utilized
to make the description compact and tractable.
On the other hand, the
second approach mainly relies on the careful analysis of the optical configurations of 
the instrument and receives feedbacks from observations to fine-tune the existing optics model.
The TinyTim software (Krist 2001) is the unique package of this type applicable to most instruments
of $HST$.

In this paper, we extend our previous efforts of the first kind (Jee et al. 2005a; 2005b; 2006; 2007) to describe the time- and 
position-dependent PSF variations of ACS/WFC now with the principal component analysis (PCA). In
our previous work, we used ``shapelets'' (Bernstein \& Jarvis 2002; Refregier 2003) to perform
orthogonal expansion of the PSFs. Shapelets are the polar eigenfunctions of two-dimensional
quantum harmonic oscillators, which form a highly localized orthogonal set.
Although the decomposition of the stars with shapelets is relatively efficient and has proven
to meet the desired accuracy for cluster weak-lensing analyses, the scheme is less than ideal in some cases. 
One important
shortcoming is that it is too localized to capture the extended features of PSFs (Jee et al. 2007; also see
\textsection\ref{section_basis_function}). 
In principle, the orthonormal
nature of shapelets should allow us to represent virtually all the features of the target image when the number of basis 
functions are sufficiently large. However, this is not a viable solution not only because the convergence is slow, but also because
the orthonormality breaks down in pixelated images for high orders as the function becomes highly oscillatory within a pixel.

The PCA technique provides us with a powerful scheme to obtain the optimal set of basis functions from the data themselves.
Unlike ``shapelets'', the basis functions derived from the PCA are by nature non-parametric, discrete, and
highly customized for the given dataset. Therefore, it is possible to summarize the multi-variate statistics,
with a significantly small number of basis functions (i.e., much smaller than the dimension of the problem).
For example, PCA has been applied to the classification of object spectra in large area surveys (Connolly et al. 1995; 
Bromley et al. 1998; Madgwick et al. 2003).
It has been shown that only a small number ($10\sim 20$) of the basis functions or $eigenspectra$
are needed to reconstruct the sample.
The application of PCA to the PSF decomposition is used by the Sloan Digital Sky Survey (SDSS) to model
the PSF variations (e.g., Lupton et al. 2001; Lauer 2002). Jarvis \& Jain (2004) used 
the PCA technique to describe the variation in the PSF pattern in the CTIO 75 square-degree survey
for cosmic shear analyses. They fit the ``rounding'' kernel component with PCA, not the PSF shape directly.
This scheme is motivated by their shear measurement technique (i.e., reconvolution to remove systematic PSF anisotropy).
However, in the current study we choose to fit the PSF shapes directly because this is more
general in the sense that the rounding kernel components are not uniquely determined for a given PSF.
In addition, our PSF library generating the PSF shapes directly has more uses in other studies.

We aim to construct a high-quality PSF library for the broadband ACS filters 
(F435W, F475W, F555W, F606W, F625W, F775W, F814W, and F850LP) 
from $>400$ archival stellar images,
which sample a wide range of the $HST$ environments (e.g., the focus values). 
Our PSF models describe ACS PSFs in rectified images, specifically, drizzled
using the Lanczos3 kernel with an output pixel scale of 0.05\arcsec (see
\textsection\ref{section_drizzling_kernel} for the justification of this choice).
The results
from this work are made publicly available on-line via the ACS team web site\footnote{The full PSF library of ACS
will become available at http://acs.pha.jhu.edu/$\sim$mkjee/acs\_psf/.}. 

We will present our works as follows.
The justification and the basic mathematical formalism of PCA are briefed in \textsection\ref{section_PCA}.
In \textsection\ref{section_application_acs}, we demonstrate how the technique can be applied to
ACS data with some test results. Focus dependency of the ACS PSFs, comparison with TinyTim, and
strategies to find matching templates are discussed in \textsection\ref{section_discussion}
before we conclude in \textsection\ref{section_conclusion}.

\section{PRINCIPAL COMPONENT ANALYSIS OF POINT SPREAD FUNCTIONS \label{section_PCA} }

\subsection{Optimal Basis Functions for Modeling a PSF Variation \label{section_basis_function}}
The concept that any vector in a vector space can be represented as a linear combination 
of the orthonormal basis vectors can be easily extended to a two-dimensional image analysis. 
The most natural set of basis vectors for a $m\times n$ resolution image is a set of
$m\times n$ unit vectors, where the $i^{th}$ unit vector represents the $i^{th}$ Cartesian coordinate axis in the
$m\times n$ dimension; the $i^{th}$ pixel value represents the amplitude along the $i^{th}$ axis. 
These $m\times n$ unit vectors form the most intuitive set of orthonormal vectors and are
in fact still a popular choice for describing a variation especially when the dimension
is low. For larger images, however, it is obvious that one needs to find alternative
basis vectors, which can describe the image features and their variations more compactly
with less number of basis vectors than $m\times n$. In the following, we will
briefly review our experiments with two potentially useful methods, namely
wavelet and shapelet decomposition schemes in an attempt to compactly model PSF features.
By discussing some fundamental limits of these two approaches, we will justify the need 
for the new scheme, PCA, to overcome these pitfalls. 

\begin{figure}
\plotone{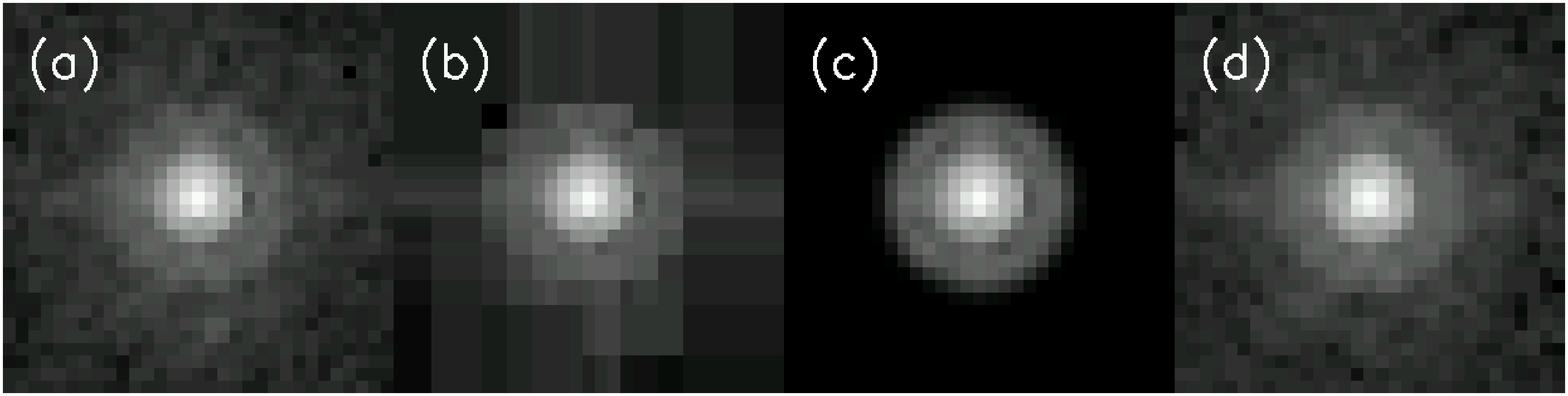}
\caption{Representation of a ACS/WFC F814W PSF with different basis functions. (a) The original $31\times31$ stellar image used
for the analyses. (b) Wavelet decomposition
with $\sim150$ $Haar$ wavelet basis functions. (c) Shapelet decomposition with 78 basis functions (shapelet order=12). 
(d) Representation with
20 basis functions that are obtained from the PCA of $\sim800$ stars.
The PSF images in (b) and (c) describe the PSF core well. However, it is obvious that many features in the PSF wing are
lost in these schemes. Although only 20 basis functions are used, the PCA method (d) captures 
many detailed features in the wing outside the second-diffraction ring, as well as the cuspiness in the PSF core.
\label{fig_psf_compare}}
\end{figure}

In astronomy, a $wavelet$ analysis has been among the most popular choices in
compressing object images, identifying objects, recognizing patterns, filtering noise, etc. 
The method has significant advantages over a traditional Fourier method particularly 
when the signal contains discontinuities and sharp spikes.
Wavelets refer to finite and fast-decaying orthogonal basis functions, 
which can efficiently represent localized signals. 
Because PSFs are in general compact, sharp, and localized, the wavelet transform can be considered as 
a tool for describing the PSF and its variation. 
However, we find that, although the scheme is very powerful in representing the
global feature, the wavelet representation with a small subset of the entire basis functions
cannot fully capture the sophisticated details of a ACS PSF.
Shown in Figure~\ref{fig_psf_compare}b is a $Haar$ wavelet representation of the
ACS/WFC PSF (Figure~\ref{fig_psf_compare}a) with $\sim16$\% (157 out of 961) of the total basis 
vectors retained. The PSF core within $\sim3$ pixel radius is satisfactory whereas most of
other features beyond $\sim3$ pixel radius are
severely smeared. Although it is possible to improve the quality of the wavelet representation
by employing more basis functions, we observe that the convergence is slow and one
has to include more than $\sim80$\% of the entire basis functions to achieve the goal.

Bernstein \& Jarvis (2002) and Refregier (2003) proposed to use shapelets
to decompose astronomical objects. Shapelets, also forming an orthogonal set,
are derived from Gaussian-weighted
Hermite polynomials, which are eigenfunctions of two-dimensional quantum harmonic
oscillators. As shapelets are based on two dimensional circular Gaussian functions, they
are somewhat more localized than $wavelets$, thus potentially more efficient in describing 
the PSF core. Figure~\ref{fig_psf_compare}c shows that
indeed the central region of the PSF is nicely recovered with 78 basis functions (shapelet order of 12); 
the second diffraction ring at $r\sim6$ pixels
is clear. It is not surprising however to observe that
the other features beyond $\sim~8$ pixels are completely washed out in 
Figure~\ref{fig_psf_compare}c because the Gaussian nature of the shapelets truncates
the profile too early to capture the apparent PSF wings. The fraction of
the flux distributed outside the second diffraction ring is less than 5 \% (compared
to the original 31$\times$31 PSF)
and
thus is negligible for some applications. In particular, if one 
looks for lensing signals in galaxy clusters, the inaccuracy in shear measurement
caused by this PSF wing truncation is overwhelmed by the shear-induced ellipticity changes.
However, in modern cosmic shear studies, the required level of systematic errors are much more
stringent, and thus we still want to develop an even better scheme that robustly
describes the PSF features on both small and large scales.

From the above two experiments, it becomes clear that any basis functions
that are derived from some analytic functions have fundamental limits in their efficiency
when we require both small and large scale features (i.e., central cuspiness and extended diffraction
pattern) of PSFs to be stringently recovered. This implies that the ideal
basis functions for a given dataset must be derived from the dataset itself. One powerful method
to achieve such a goal is PCA. Also known as Karhunen-Loeve transformation (KLT), PCA
provides a method for obtaining optimal basis functions highly tailored to a given problem.
As will be briefly summarized in \textsection\ref{section_PCA_formalism}, PCA allows us
to keep the subset of basis functions 
that has the largest variance. The principal
components (hereafter we use the terms, principal component, basis function, and eigen-PSF interchangeably)
with the lowest variances are dominated by noise and can be safely 
discarded to reduce the dimension
of the problem.

\begin{figure}
\plotone{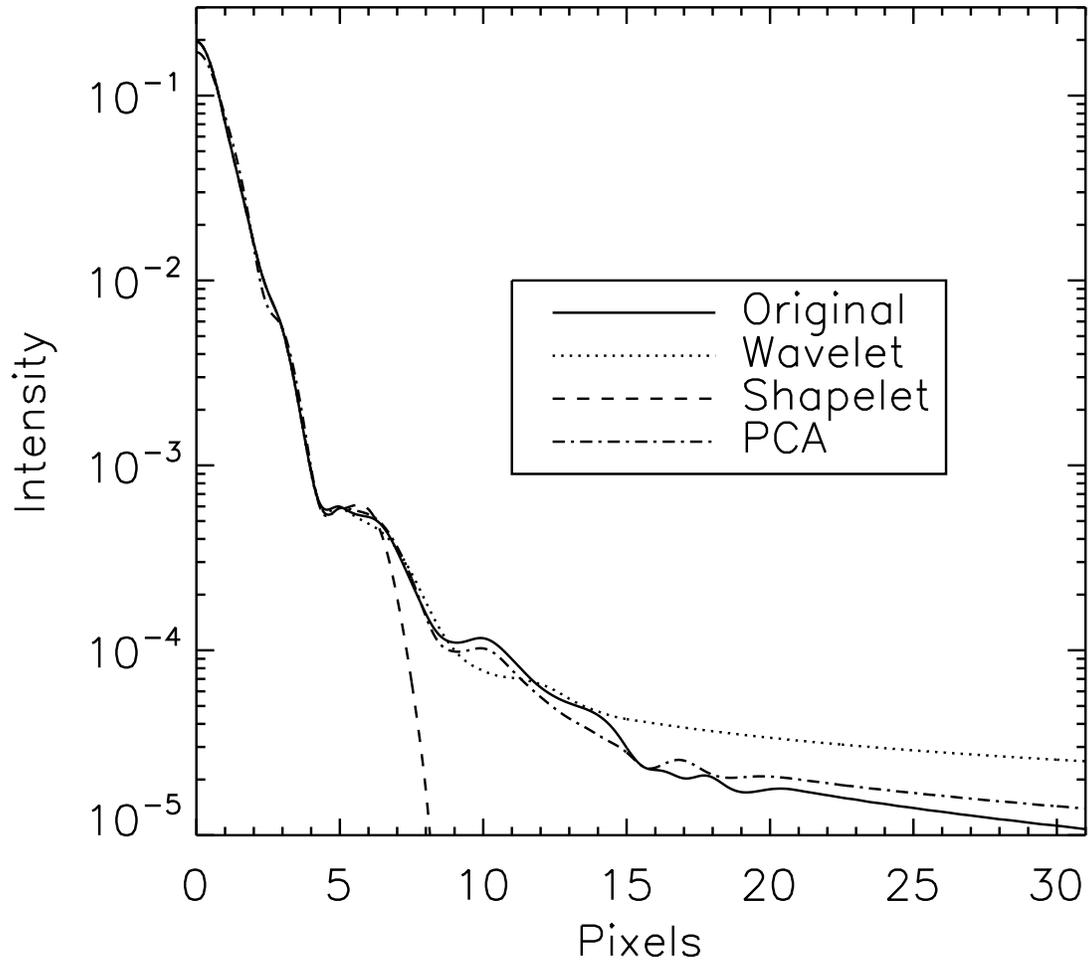}
\caption{Comparison of radial profiles in different PSF representation. The PCA representation of the PSF has
the radial profile closest to that of the original. The shapelet method generates the PSF that truncates
at $r\simeq8$ pixels. The representation with 150 $Haar$ wavelets appears to approximate the radial profile
of the original closely, but we see in Figure~\ref{fig_psf_compare} that the two-dimensional representation
is unacceptable.
\label{fig_psf_profile}}
\end{figure}

We display the PSF image constructed with the first 20 principal components in Figure~\ref{fig_psf_compare}d.
The 20 principal components are obtained by analyzing $\sim870$ stars in the same exposure. 
The dramatic
improvement in the recovery of the original PSF is apparent not only in the core, but also in the diffraction
pattern far from  the core. 
This is again verified in the comparison of the radial profiles in different
representation of the PSF (Figure~\ref{fig_psf_profile}). The PCA method gives
the radial profile closest to that of the original (we note that the PCA method slightly 
fits noise at $r>8\arcsec$ because the signal outside the second diffraction ring is very weak. Potentially
one can improve the sampling by including the wings of saturated stars as is done by Anderson \& King [2006]).
The shapelet method generates the PSF that truncates
at $r\simeq8$ pixels. The representation with 150 $Haar$ wavelets appears to approximate the radial profile
of the original closely, but we see in Figure~\ref{fig_psf_compare} that the two-dimensional representation
is unacceptable. 
Therefore, considering both the relatively small number of basis functions and the quality
of the reproduction, we choose the PCA approach for our subsequent analysis of the time- and position-dependent PSF of WFC.

\subsection{Mathematical Formalism \label{section_PCA_formalism} }

Imagine that we have a dataset consisting of $N$ observations (e.g., stars), each with $M$ observable
properties (e.g., pixel values).
If the $M$ observable properties do not change greatly between observations,
the dataset forms a cloud of $N$ points in an $M$-dimensional space.
We want to construct a set of
$P$ ($\ll Min\{N,M\}$) orthonormal vectors that describes the subspace in
the following manner.
\begin{itemize}
\item The first significant vector is defined as the axis with
a minimal mean distance from each point. 
\item The second significant vector is orthogonal to the first significant vector and 
minimizes the mean distance from each point. 
\item The $P^{th}$ significant vector
is orthogonal to the previous $(P-1)$ significant vectors and minimizes the mean distance.
\end{itemize}
\noindent
These $P$ significant vectors are called principal components (PC) of the system.

One way to construct such a new orthonormal basis is Singular Value Decomposition (SVD; Press et al. 1992). 
We can express the above dataset as a $N \times M$ matrix $\bvec{S}$. According to
the SVD theory, any $N \times M$ matrix whose number of rows $N$ is greater than or equal to
its number of columns $M$, can be rewritten as the product of an $N \times M$ column-orthogonal
matrix $\bvec{U}$, an $M \times M$ diagonal matrix $\bvec{W}$ with positive or zero elements, and the transpose
of an $M \times M$ orthogonal matrix $\bvec{V}$

\begin{equation}
\bvec{S}=\bvec{UWV}^{T}
\end{equation}

It is easy to show that once the matrix $\bvec{S}$ is expanded in the way above, any $ij^{th}$ element of the matrix
$\bvec{S}$
can be reconstructed by
\begin{equation}
S_{ij} = \sum_{k=1}^{M} w_k U_{ik} V_{jk}. \label{eqn_svd}
\end{equation}
\noindent
Equation~\ref{eqn_svd} helps us to realize that, if some of the singular values $w_k$ (elements of $\bvec{W}$) 
are tiny, we can approximate the matrix $\bvec{S}$ by replacing those small $w_k$'s with zeros.
This effectively reduces the number of columns in $\bvec{U}$ and $\bvec{V}$. The remaining columns of $\bvec{V}$
serve as the principal components that form an orthonormal basis.

A geometric meaning of these principal components is that they define the principal axes of the error
ellipsoid. Consequently, they are eigenvectors of the covariance matrix of the dataset $\bvec{C}$, diagonalizing
$\bvec{C}$ with eigenvalues of $w_i$, which
also illustrates that PCA is the rotation of the system into a new basis (i.e., principal components) to
describe the data in terms of statistically independent quantities. 

In practice, PCA necessitates preprocessing of the data typically involving mean subtraction and
normalization. The exact procedure highly depends on the statistical nature of the problem and we will
discuss the issues in \textsection\ref{section_implementation}.

\section{PCA APPLICATION TO ACS DATA \label{section_application_acs} }

\subsection{Implementation \label{section_implementation}}
In the current section, we demonstrate how we can describe the PSF variation 
observed in a single ACS/WFC exposure with PCA described in \textsection\ref{section_PCA_formalism}.
We select the F814W observation of the 47 Tuc field taken on 6 May 2002 in two 30s exposures
(dataset ID = J8C0D1051).
The image was part of a series of observations to derive flat-fielding model of the instrument (PROP ID 9018).
The low level CCD processing was carried out using the STScI standard ACS calibration
pipeline (CALACS; Hack et al. 2003). In correcting the geometric distortion, 
we used a Lanczos3 drizzling kernel with an output pixel size of 0.05$\arcsec$.
In our previous analyses (Jee et al. 2005a; 2005b; 2006; 2007), this combination of drizzling parameters 
has been verified to be among the optimal choices in minimizing the aliasing, the noise
correlation, and the broadening of the PSF. We discuss some important differences arising from
different choices of drizzling parameters in \textsection\ref{section_drizzling_kernel}.

The field is moderately crowded (Figure~\ref{fig_starfield_image}) and 
we were able to select $\sim 870$ bright ($m_{VEGAMAG}\lesssim 13.3$), unsaturated, and isolated 
(no adjacent stars within a $\sim20$ pixel radius) stars.
After creating a postage stamp image ($31$ pixel $\times31$ pixel) for each star, we applied sub-pixel shifts 
so that the peak always lies on the center of a pixel. Omitting
this procedure would result in the variance of the system largely dominated by the location of the peaks within
pixels. The sub-pixel shifts were carried out with bicubic interpolation, which closely approximates
the theoretically optimal, windowed sinc interpolation by cubic polynomials. We find that although
bicubic interpolation slightly softens PSF cores relative to the results from windowed sinc interpolation,
the latter creates more frequent other types of artifacts such as occasional 
negative pixels (in theory, the sinc interpolant is valid for a Nyquist-sampled image).

We need to express the PSF images with one-dimensional vectors. Because the modeling size is
$31\times 31$, each vector has 961 elements. Our matrix $\bvec{S}$ describing the dataset
has $M=961$ columns and $N=870$ rows. There are more columns than rows, and 
the SVD above will yield $M-N$ (or more because of the degeneracies) zero or negligible $w_j$'s.
However, the remaining $\bvec{V}$ still contains useful principal components of the system, which can efficiently
represent the sample.

We normalized $\bvec{S}$ in such a way that the sum of the elements in each row is unity 
after subtracting the background value (flux normalization).
Next, we created a mean PSF by taking averages along the columns. Then, we subtracted this mean PSF from each row.
The resulting matrix $\bvec{S}$ consists of deviations from this mean PSF.

We perform SVD of $\bvec{S}$ by diagonalizing the covariance matrix $\bvec{C}$, which
is the outer product of $\bvec{S}$ with itself. 
The resulting eigenvectors and the eigenvalues are the PCs and the variances
of the matrix, respectively. Finally, we sort the result in order of decreasing variances.
Figure~\ref{fig_variance} illustrates that the first $\sim20$ PCs account for more than 90\% 
of the total variance.
Each of the remaining 900 principal components is responsible for less than 1\% of the total, likely to be
associated with noise rather than to contain the real signal.

We determined the PC coefficients (i.e., amplitudes along the eigenvectors) down to the 20th 
largest component by multiplying $\bvec{S}$ to the eigenvectors for the selected stars.
The spatial variation of each coefficient is fit with the following
polynomial:
\begin{equation}
P_{i}  = a_{00} + a_{10} x + a_{01} y + a_{20} x^2 + a_{11} xy + a_{02} y^2 +  \cdot \cdot \cdot .
\end{equation}
\noindent
We found that a fifth order in $x^i y^j$ (i.e. $i+j \le 5$) is sufficient to describe the pattern and higher
order polynomials do not improve (sometimes worsen) the agreement between model and data.
The total number of coefficients
necessary to model the PSF variation for a entire WFC frame is $15\times20=300$.

\subsection{Test Results \label{section_test_result}}

There might exist a number of ways to 
compare our PSF model obtained in \textsection\ref{section_implementation} 
with the real PSFs, depending on how one chooses to characterizes PSFs. In this paper, we characterize
PSFs by their ellipticity and width because these parameters are natively related to the systematics in
weak-lensing measurements and also are sensitive to the charge diffusion and the local focus offset.

We measure a star's ellipticity and width using the following quadrupole moments,

\begin{equation}
Q_{ij} = \frac{ \int d^2 \theta W(\symvec{\theta}) I(\symvec{\theta}) (\theta_i - \bar{\theta_i})(\theta_j - \bar{\theta_j}) }
           {\int d^2 \theta W(\symvec{\theta}) I(\symvec{\theta}) }, \label{eqn_quadrupole}
\end{equation}	   
\noindent
where $I(\symvec{\theta})$ is the pixel intensity at $\symvec{\theta}$, $\bar{\theta}_{i(j)}$ is
the center of the star, and $W(\symvec{\theta})$ is the weight function required to suppress the noise in the outskirts
(we choose a Gaussian with a FWHM of 2 pixels throughout the paper).
With Equation~\ref{eqn_quadrupole} at hand, it is now possible to define the star's ellipticity
in the following two ways:
\begin{equation}
\symvec{\delta} = \left ( \frac{Q_{11}-Q_{22}} {Q_{11}+Q_{22}} , \frac {Q_{12}} {Q_{11}+Q_{22}} \right )
\end{equation}
and
\begin{equation}
\symvec{\epsilon} = \left ( \frac{Q_{11}-Q_{22}} {Q_{11}+Q_{22}+2(Q_{11} Q_{22}-Q^2_{12})^{1/2}} , 
\frac {Q_{12}} {Q_{11}+Q_{22}+2(Q_{11} Q_{22}-Q^2_{12})^{1/2}} \right ) \label{eqn_delta}
\end{equation}
\noindent
For an ellipse with axis ratio $r$, $|\symvec{\delta}|$ and $|\symvec{\epsilon}|$ 
correspond to $(1-r^2)/(1+r^2)$ and $(1-r)/(1+r)$, respectively.
In the current paper, we select Equation~\ref{eqn_delta} as our definition of ellipticity, referring to
the first and second components of $\symvec{\epsilon}$ as $\epsilon_{+}$ and $\epsilon_{\times}$, respectively. 
The size of a star can be similarly
defined using the above quadrupole moments $Q_{ij}$. One common choice is
\begin{equation}
b=\sqrt{Q_{11}+Q_{22}} \label{eqn_psf_width},
\end{equation}
which we adopt in this work.

\begin{figure}
\plotone{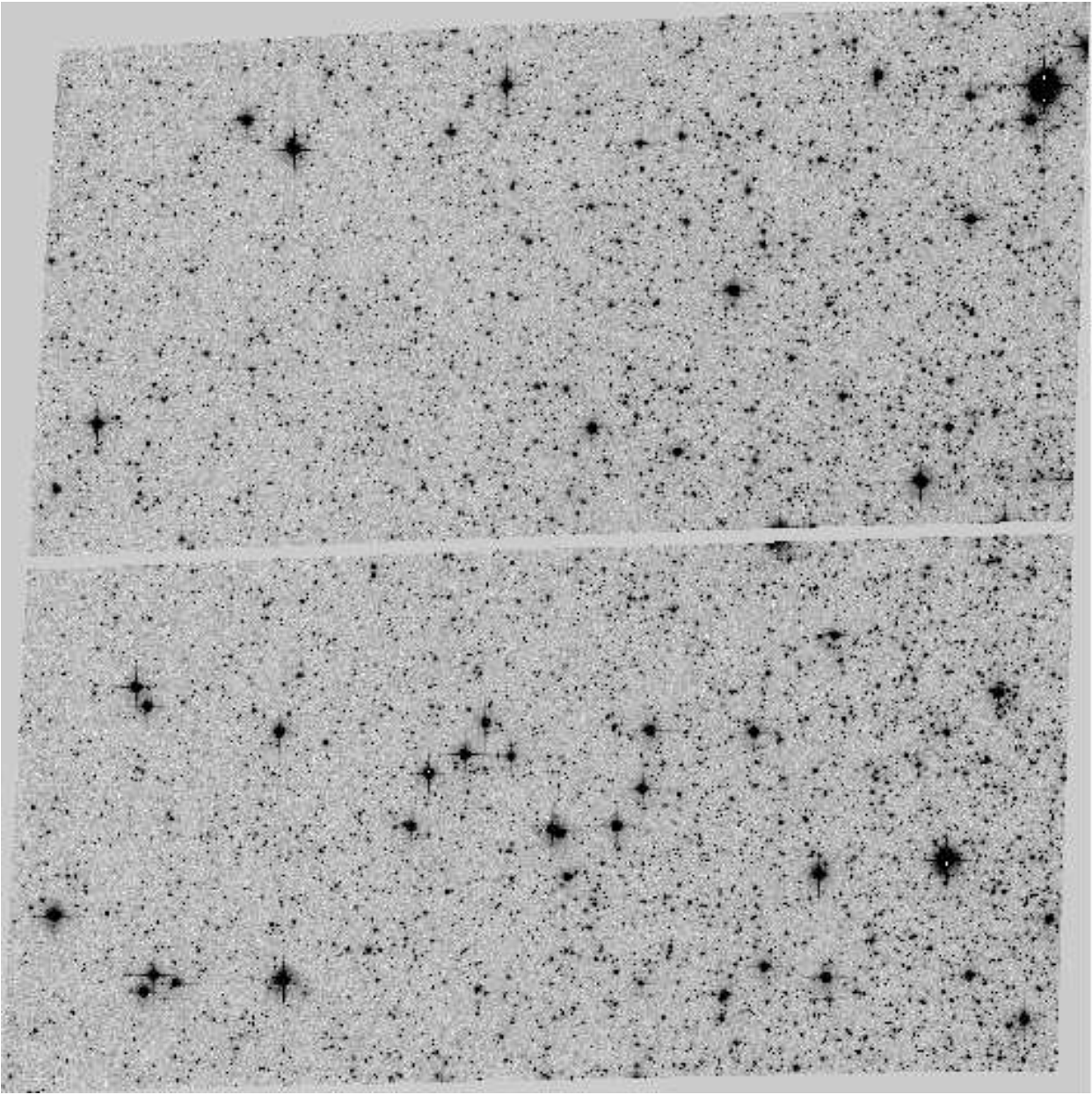}
\caption{WFC observation of the moderately crowded field of the globular cluster 47 Tuc. The image was taken in F814W on 19 April 2002. We choose
this particular dataset to demonstrate how we can model the PSF variation with PCA. 
\label{fig_starfield_image}}
\end{figure}

\begin{figure}
\plotone{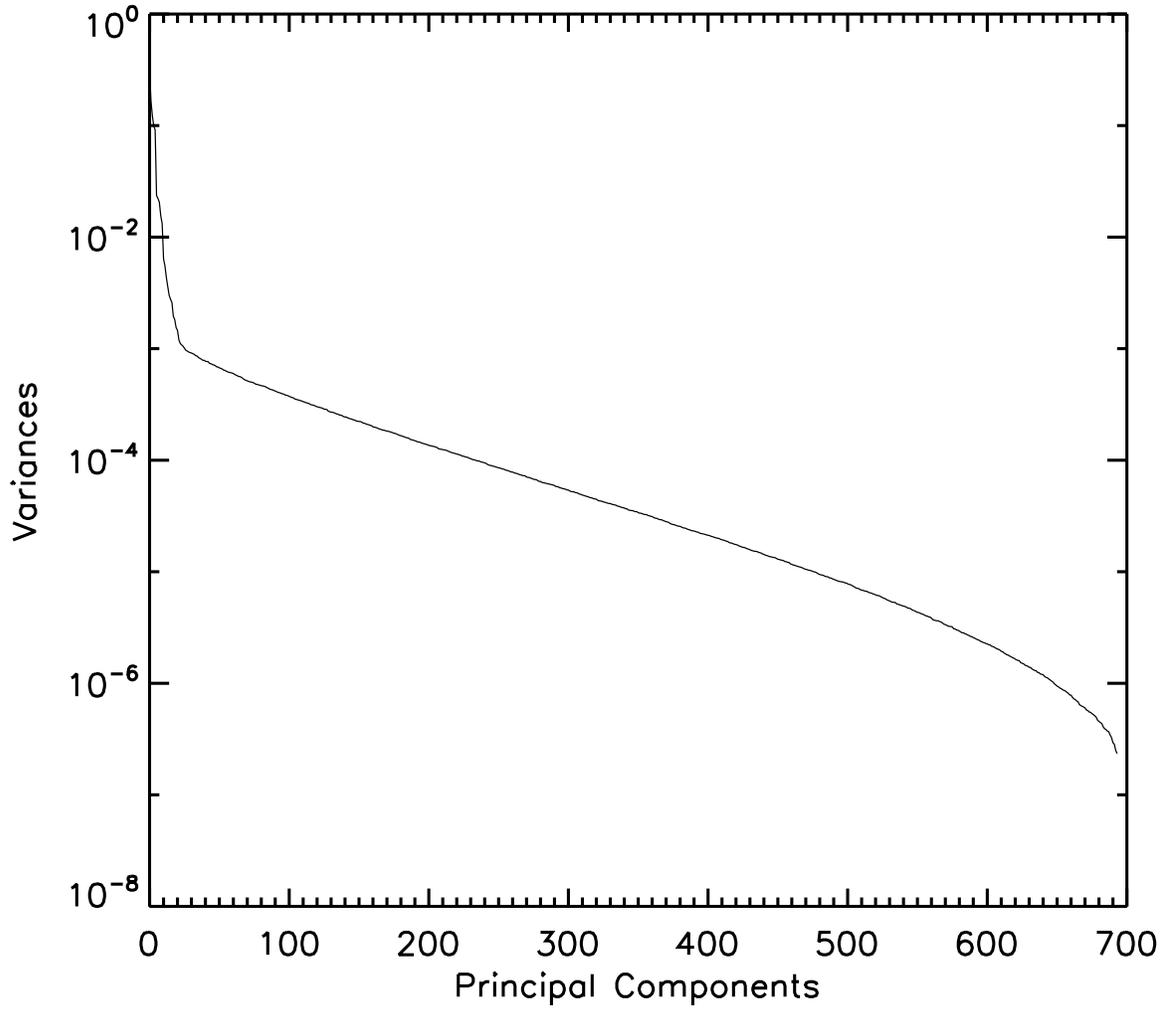}
\caption{Variances of principal components. Principal components are rearranged in orders of decreasing variances. The
first $\sim20$ principal components dominantly contribute to the total variance.
\label{fig_variance}}
\end{figure}

\begin{figure}
\plotone{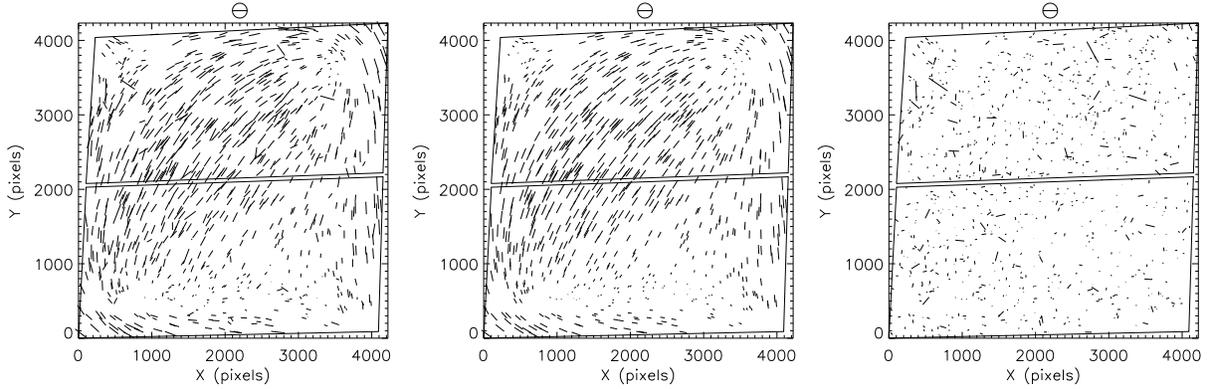}
\caption{Position-dependent WFC PSF ellipticity variation. In the left panel, we display the ellipticity of the stars directly measured
from the image (Figure~\ref{fig_starfield_image}). The size and the orientation of the ``whiskers" represent the magnitude of ellipticity and
the direction of elongation, respectively. Our PCA model nicely reproduces the position-dependent ellipticity variation (middle panel).
The residuals (right panel) between the PCA model and the direct measurements are very small. It is apparent that our PSF model obtained through PCA stringently recovers the observed ellipticities.
With 3 $\sigma$ outliers discarded, the mean absolute deviation $<|\delta \symvec{\epsilon}|>$ is $(6.5\pm0.1)\times 10^{-3}$, and the mean ellipticity 
$< \delta \symvec{\epsilon} >$ is $[ (1.1\pm2.2)\times10^{-4} , (2.3\pm1.4)\times10^{-4}]$.
\label{fig_ellipticity_recovery}}
\end{figure}

In the left panel of Figure~\ref{fig_ellipticity_recovery}, we display the ellipticities of the $\sim870$ isolated stars found
in Figure~\ref{fig_starfield_image}. The size and the orientation of the ``whiskers" represent the magnitude of ellipticity and
the direction of elongation, respectively. The majority of the stars are stretched approximately parallel to the $y=x$ line.
This direction is roughly tangential to the vector pointing towards the telescope axis, and this type of pattern is observed when 
$HST$ is at its nominal
``negative'' focus (the actual focus offsets on the surface of the WFC detector can be positive in certain regions 
because of the curvature of the focal plane and the detector height variation). We repeat this ellipticity measurement
for our model PSFs and the results are shown in the middle panel of Figure~\ref{fig_ellipticity_recovery}, which
displays the predicted ellipticities of PSFs at the same star positions. We plot the residual ellipticities in
the right panel. It is apparent that our PSF model obtained through PCA robustly recovers the observed ellipticities.
With 3 $\sigma$ outliers discarded, the mean absolute deviation $<|\delta \symvec{\epsilon}|>$ is $(6.5\pm0.1)\times 10^{-3}$, 
and the mean ellipticity 
$< \delta \symvec{\epsilon} >$ is $[ (1.1\pm2.2)\times10^{-4} , (2.3\pm1.4)\times10^{-4}]$.

\begin{figure}
\plotone{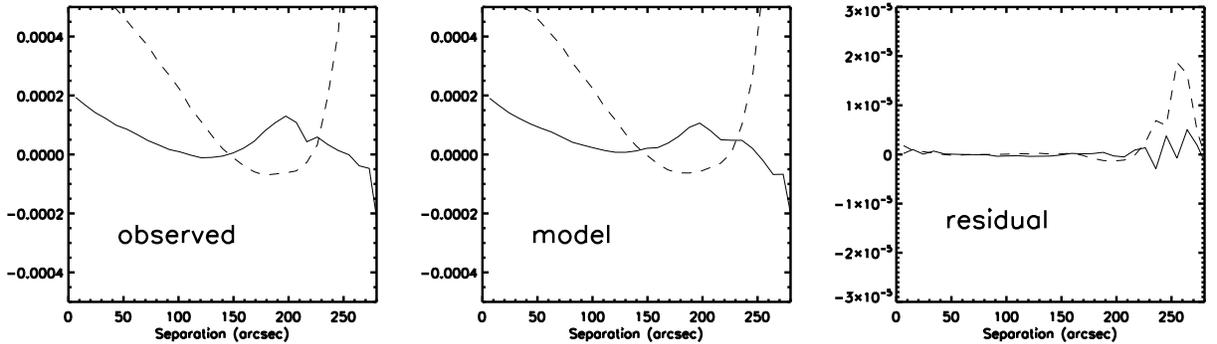}
\caption{Ellipticity correlation functions for the observed PSF (left), the model (middle), and the residual ellipticities (right).
The solid and dash lines represent $\xi_{+}$ and $\xi_{\times}$, respectively.
The amplitude of the residual ellipticity correlation is $\sim10^{-7}$ (after discarding the 
values at $\theta > 220\arcsec$, which are spuriously high due to the poor statistics in this regime
and an artifact of the interpolation), approximately three orders of magnitude
lower than the uncorrected values in the other panels. We do not display error bars to avoid clutter. The typical
size of error bars at $\theta < 220\arcsec$ is $\sim10^{-7}$.
\label{fig_e_corr}}
\end{figure}

Another way to quantify the quality of the ellipticity representation of a PSF model is to investigate
the ellipticity correlation as a function of separation $\theta$:
\begin{equation}
\xi_{+} (\theta) = < \epsilon_{+} (r) \epsilon_{+} (r+\theta) >
\end{equation}
and
\begin{equation}
\xi_{\times} (\theta) = < \epsilon_{\times} (r) \epsilon_{\times} (r+\theta) >.
\end{equation}
We show in Figure~\ref{fig_e_corr} the ellipticity correlation functions for the
the observed PSF (left), the model (middle), and the residual (right).
The solid and dash lines represent $\xi_{+}$ and $\xi_{\times}$, respectively.
The amplitude of the residual ellipticity correlation is $\sim10^{-7}$ (after discarding the 
values at $\theta > 220\arcsec$, which are spuriously high due to the poor statistics in this regime
and an artifact of the interpolation), approximately three orders of magnitude
lower than the uncorrected values.

\begin{figure}
\plotone{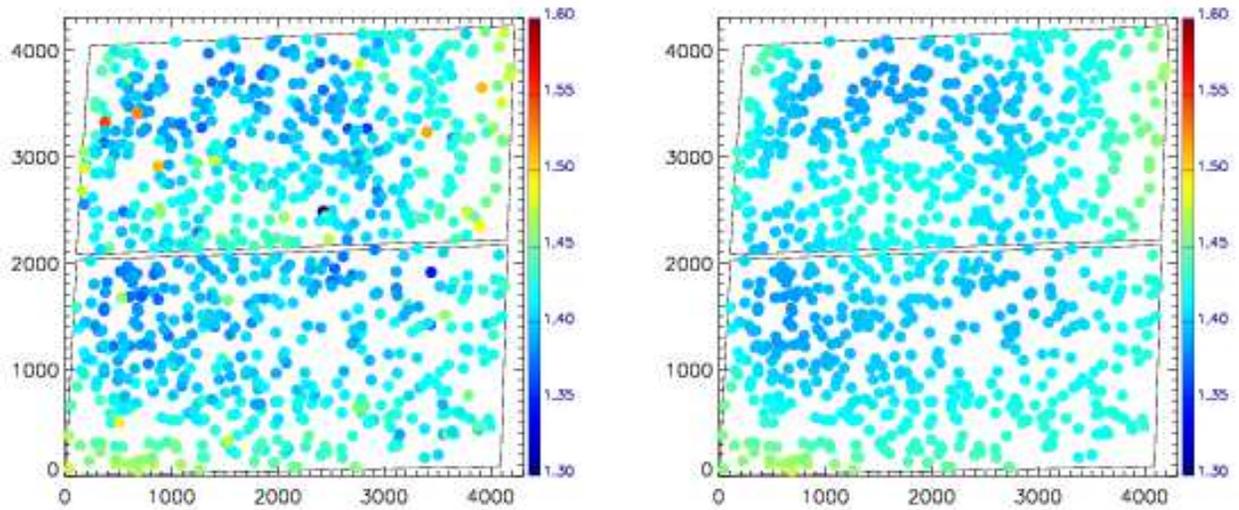}
\caption{Position-dependent WFC PSF width variation. The left panel shows our direct measurement of the PSF widths from the stars
in Figure~\ref{fig_starfield_image}. Our PSF model with PCA closely reproduces this observed PSF width variation (right).
The detection of the ``hill'' at $x\sim1500$ and $y\sim2200$ and the ``moat'' surrounding the hill is seen in both panels
though the ``hill'' looks slightly less pronounced in the right panel because of the employed polynomial interpolation
The stars in the hill in the right panel are $\sim0.4$\% smaller whereas the stars in the moat are $\sim0.3$\% larger.
\label{fig_psf_width}}
\end{figure}

A size of the PSFs (eqn.~\ref{eqn_psf_width}) is also a useful quantify in characterizing PSFs. In general,
both the aberration-induced elongation and the charge diffusion are responsible for the broadening
of WFC PSFs. Krist (2003) noticed that the PSF width variation by charge diffusion 
remarkably resembles the pattern of the WFC CCD thickness variation. The blurring is most severe in the
central region where the CCD layer is the thickest, and the $r\sim 100 \arcsec$ annulus surrounding this 
region has the least charge diffusion, consistent with its lowest thickness.
The left panel
of Figure~\ref{fig_psf_width} shows our estimation of the position-dependent WFC PSF width variation measured from the stars
in Figure~\ref{fig_starfield_image}. The global pattern nicely agrees with the result of Krist (2003) (i.e.,
the detection of the ``hill'' at $x\sim1500$ and $y\sim2200$ and the ``moat'' surrounding the hill).
Because the image is taken in F814W, the charge diffusion effect is somewhat reduced (compare this with Figure
2 of Krist 2003 showing the variation in F550M). In addition, we note that the PSF widths are
greatest along the field boundary, which is by and large due to the optically induced PSF elongation.
We display the PSF width variation predicted from our model in the right panel of Figure~\ref{fig_psf_width}.
The employed polynomial interpolation smooths the variation and slightly flattens the ``hill'' and ``moat" features.
The stars in the hill in the right panel are $\sim0.4$\% smaller whereas the stars in the moat are $\sim0.3$\% larger.

Krist (2003) claimed that there was a height difference of $\sim0.02\mu \mbox{m}$ between the two CCDs, which
manifested itself as a discontinuity of the PSF pattern across the gap. The observed
PSF size variation (left panel of Figure~\ref{fig_psf_width}) seems to show faint indications
of the height difference. However, we do not
find any noticeable discontinuity in the PSF ellipticity pattern across the gap in Figure~\ref{fig_ellipticity_recovery}.

\section{DISCUSSION \label{section_discussion}}

\subsection{Effects of Drizzling Methods on PSF Characteristics \label{section_drizzling_kernel}}
Aliasing occurs when a signal that is continuous in space is sampled with finite resolution. 
PSF shapes from rectified ACS images suffer this aliasing twice, first when 
photons are collected in the discrete CCD grid, and second when the raw data are remeshed
for the geometric distortion correction. Dithering mainly helps to reduce the first
aliasing by changing the sub-pixel position of the PSF centers within a pixel, effectively
increasing the sampling resolution of the detector beyond its physical pixel size.
The second aliasing arising from the input and output pixel offsets is mitigated by carefully
selecting an interpolation scheme.
Here, we focus on the second issue: the relation between interpolation scheme (i.e.,
parameters set in drizzling) and observed PSF characteristics.

Although quite a few combinations of drizzling kernels, output pixel sizes, and drop sizes are possible, 
we consider the following three cases:
\begin{itemize}
\item Lanczos3 kernel, $0.05\arcsec$ output pixel, and pixfrac$=1$,
\item Square kernel, $0.05\arcsec$ output pixel, and pixfrac$=1$,
and
\item Gaussian Kernel, $0.03\arcsec$ output pixel, and pixfrac$=0.8$.
\end{itemize}
The first case is of course the choice in the current paper. The second case is selected because it is the
default setting in the STScI pipeline and is most frequently used. The last one is favored by
Rhodes et al. (2007), who argued that this combination gave the minimal aliasing in their experiments.

\begin{figure}
\plotone{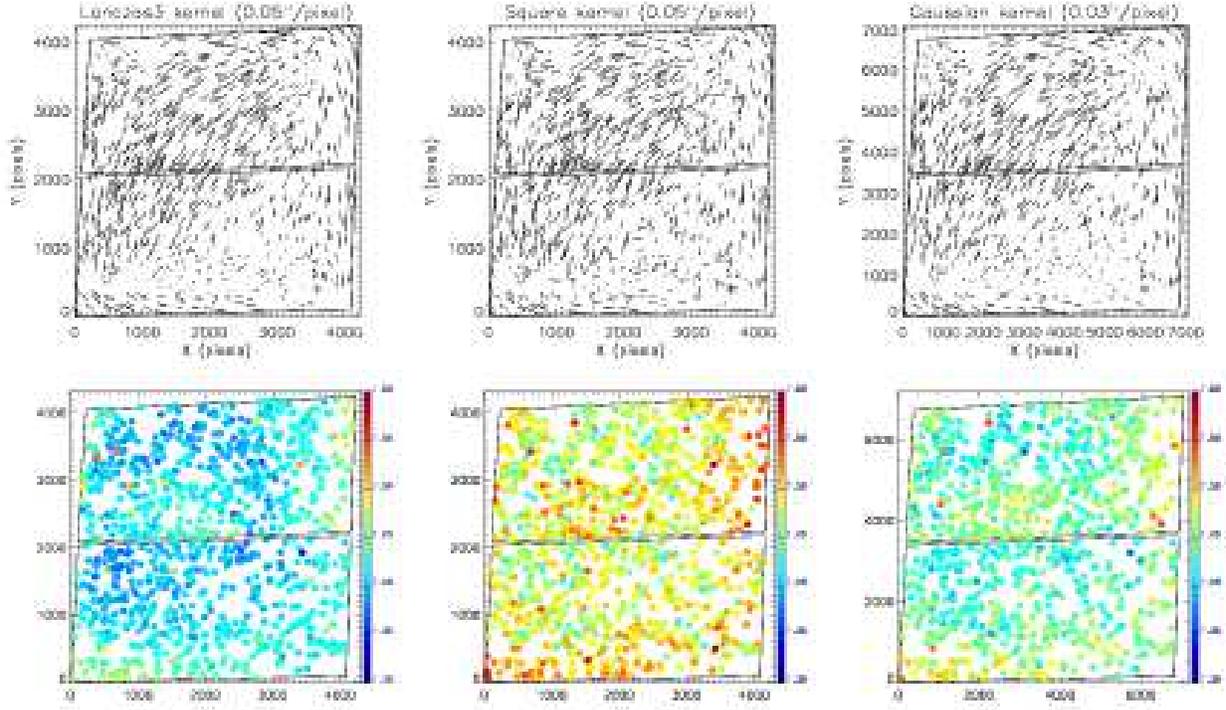}
\caption{Effects of drizzling kernel and output pixel size on observed PSFs.
Top and bottom panels show PSF ellipticity and width variation,
respectively, for different drizzling methods. The case for the Lanczos3 kernel ($left$) already shown in Figure~\ref{fig_ellipticity_recovery}
and Figure~\ref{fig_psf_width} is reproduced here to ease the comparison.
Aliasing is non-negligible for the case of the square kernel ($middle$) used in conjunction with an output pixel size of 0.05$\arcsec$.
In addition, it is obvious that this choice of drizzling method also broadens the observed PSFs most severely.
Drizzling with a Gaussian kernel with a pixel scale of 0.03$\arcsec$ ($right$) reduces the aliasing in the ellipticity
measurements seen in the square kernel image. However, this level of the aliasing reduction is already achieved in
the case of the Lanczos3 kernel. 
The observed PSF widths in the Gaussian kernel is smaller than the PSF widths in the case of the square kernel, but larger
than the PSF widths in the case of the Lanczos3 kernel (note that we multiplied $0.03/0.05=0.6$ to the PSF width measurements from the Gaussian kernel 
image to remove the difference arising from the pixel scale discrepancy).
\label{fig_kernel}}
\end{figure}

We compare the results from these three methods by examining the PSF ellipticity and width distribution across the
WFC detector (Figure~\ref{fig_kernel}). As noted by Rhodes et al. (2007), it is obvious that drizzling with 
square kernel and 0.05$\arcsec$ output pixel produces the most severe aliasing among the three. 
The residuals between the observed (top middle) and the PCA interpolated stars 
(as calculated in the second panel of Figure~\ref{fig_ellipticity_recovery})
have a
mean absolute deviation of $<|\delta \symvec{\epsilon}|>=(1.29\pm0.03)\times 10^{-2}$
with a center at $< \delta \symvec{\epsilon} >=[ (3.5\pm4.9)\times10^{-4} , (4.7\pm1.2)\times10^{-4}]$.
In addition, the PSF blurring is also the largest ($bottom$ $middle$) in the square-kernel-drizzled image.
The mean PSF width is $1.486\pm0.001$, about 5.4\% larger than the value we obtain
from the Lanczos3-kernel-drizzled image ($1.410\pm0.001$).

Drizzling with a Gaussian kernel with a pixel scale of 0.03$\arcsec$ ($top$ $right$) 
reduces the aliasing in the ellipticity measurements compared to that in the square kernel image. 
The residuals have a mean absolute deviation of 
$<|\delta \symvec{\epsilon}|>=(6.9\pm0.1)\times10^{-3}$ with a center at
$< \delta \symvec{\epsilon} >=[ (2.5\pm2.3)\times10^{-4} , (4.8\pm1.5)\times10^{-4}]$.
However, the mean absolute deviation is $\sim6$\% higher than in the Lanczos3 case.
Also, although it is true that the PSF broadening is mitigated compared to
the square-kernel drizzling, the mean PSF width ($1.445\pm0.001$) is $\sim3$\% larger
than the Lanczos3-kernel PSF width.

Based on the above experiment, we claim that the Lanczos3 kernel with an output pixel size of 0.05 $\arcsec$ should be a preferred choice
in weak-lensing analyses (and also perhaps in other analyses that require sharpest images). 
Although the choice of the Gaussian kernel with a pixel scale of 0.03\arcsec provides
a competitive performance 
in terms of the reduction of aliasing and the sharpness of PSF,
we have observed that noise correlation is the most severe in this case (this pitfall is
also noted by Rhodes et al. [2007]). Noise correlation between adjacent pixels creates
visible moir{\'{e}} patterns in the image.

However, we comment that 
the Lanczos3 kernel occasionally
produces some cosmetic artifacts in the region where flux gradients change abruptly (e.g., centers of saturated stars,
wings of bright stars, missing data points, etc.). Nevertheless, these occasional cosmetic artifacts in
individual stars are not of concern in the current PSF sampling because they are efficiently filtered out through PCA.

\subsection{Focus Dependency}
In \textsection\ref{section_test_result}, we studied how the ellipticity and the size of WFC PSFs
vary across the field for the particular dataset (the F814W filter on 19 April 2002). 
If the pattern remained the same throughout the life of $HST$ or if the change were negligible, the
issue of correcting PSF effects would be trivial. Unfortunately, an observed PSF pattern
is not stable, but changes over time, depending largely on the focus status of $HST$.
The main cause of the $HST$ focus variation is the combination of both the constant shrinkage and 
the thermal breathing of the optical telescope assembly (OTA) truss structure. Occasional adjustments of the secondary mirror position
were applied (e.g., on 24 December 2004) to compensate for the former long-term change. The latter thermal
breathing occurs as the OTA truss structure 
expands during Earth occultation and contracts after the occultation. The typical amplitude
of the focus change during an orbit is $3\sim5 \mu m$. This small focus variation does not
severely affect the quality of ACS observations in general. However, it produces
conspicuous changes in the PSF ellipticity and width variation across the detector.

In Figure~\ref{fig_many_pattern} and \ref{fig_many_pattern_width}, we show the time-dependent PSF pattern
in ellipticity and width, respectively, observed in 30 different
F435W exposures. In both figures, time increases to the right and to the bottom; the observation date and time are denoted
in the $year-month-day$ and $hour-minute-second$ (UT) format above each panel.
The exposures are not homogeneously sampled in time
(e.g., the first 9 exposures are taken on the same observation date). It is clear that the PSF ellipticity and width patterns
vary quite significantly. When the instrument is at negative focus, the 
``whiskers'' are on average elongated from lower left to upper right as already seen for the case in 
\textsection\ref{section_test_result}.
We observe that these negative focus
patterns dominate over positive focus patterns 
not only in F435W, but also in other filter observations. At positive focus, the whiskers are approximately perpendicular to
the pattern observed at negative focus (e.g., plots in the fourth row of the second and the sixth columns).
Comparison between Figure~\ref{fig_many_pattern} and \ref{fig_many_pattern_width} show that
the average PSF widths per pointing are in general
proportional to the average magnitudes of ellipticities (i.e., size of whiskers).
Moreover, we realize that the PSF width variation pattern is potentially a more sensitive
measure of the HST focus and helps us to characterize the pattern more precisely.
For example, the first 9 exposures taken on the same observation date appear to
possess ellipticity patterns very similar to one another. If one is somehow asked to
select two PSF patterns that were observed under similar circumstances, the task based
on the visual inspection of these ellipticity plots is quite challenging. However,
with the aid of Figure~\ref{fig_many_pattern_width} one can easily tell
that the first one can pair with the fourth one, the second one with the third one, 
the sixth one with the seventh one, etc.

A close examination of the time-dependent variation of the pattern suggests that the patterns are 
repeatable even if their observation epochs are quite apart.
For example, the two observations taken on 24 October 2002 (2nd row and 5th column) and 6 September 2003 (3rd row and 6th
column) match each other not only in the ellipticity pattern, but also in the PSF width pattern.
Although this repeatability does not necessarily guarantee that the PSF pattern is uniquely determined by
a single parameter (focus), it provides important justification that we can apply the PSF templates
obtained from these stellar fields to science observations taken at different epochs.

We suspect that other factors such as velocity aberration, detector plane tilt, pointing accuracy degradation, etc.
also might modulate the observed pattern. As suggested by Jarvis \& Jain (2006), PCA of
the polynomial coefficients may help us to determine the number of degrees of freedom in future
investigations.

\subsection{Comparison with a shapelet approach}
In our previous studies, we used shapelets to interpolate PSF variations across WFC. Although shapelets
provide competitive efficiency in describing ACS PSFs, their performance is somewhat inferior to the 
current method obviously because the basis functions derived from PCA is optimally customized to the
given data. In the following, we present quantitative comparison using the same dataset (J8C0D1051)
analyzed with PCA in \textsection\ref{section_implementation}.

We choose a shapelet order to be eight and apply a 4th order polynomial interpolation. 
As stated in Jee et al. (2005a), increasing the order of polynomials beyond the third order does
not improve the fit. 
A shapelet order of eight contains 45 independent basis functions. We again emphasize that
increasing the order of shapelet beyond this at the expense of computation time does not noticeably
improve the quality of the ACS PSF representation (sometimes this makes the interpolation unstable
as the high order terms start fitting the noise).

The left panel of Figure~\ref{fig_shapelet_performance} shows the ellipticity residuals between the observed
stars and the shapelet model. The mean absolute deviation $<|\delta \symvec{\epsilon}|>$ is 
$(6.8\pm0.1)\times 10^{-3}$ after 3 $\sigma$ outlier rejection. This is very close to the value
that we obtained from the PCA method ($(6.5\pm0.1)\times 10^{-3}$). However, the residual ellipticity correlation
(middle panel of Figure~\ref{fig_shapelet_performance}) illustrates that the systematic errors of
the shapelet model (black) is somewhat higher than the PCA model (red); we see higher
correlation on small scales ($\lesssim 100\arcsec$)
and higher anti-correlation on large scales ($\gtrsim 150\arcsec$). 

In the right panel of Figure~\ref{fig_shapelet_performance}, we display the ACS PSF width variation
predicted by the shapelet model. Comparison of this plot with Figure~\ref{fig_psf_width}
shows that the shapelet model dampens the variation pattern. The ``moat'' and ``hill''
features are hard to identify although the model satisfactorily describes the broadening at the
field boundaries (the lower-right and upper-right corners). We note that the widths of ``hill'' stars
here are $\sim4$\% smaller than those of the observed stars.

In summary, the ACS PSF variation model through shapelet coefficient interpolation performs
competitively well in representing the ellipticities although we note that the systematics in
the shapelet PSF model is higher than in the PCA PSF model. The shapelet PSF model does not
fully describe the ACS PSF width variation, under-representing the ``moat'' and ``hill' features.
These shortcomings are not worrisome in typical cluster weak-lensing
analyses, where the residual systematics are still far smaller than galaxy shape or foreground-contamination noise.
However, the under-representation of the PSF widths makes the shapelet model inadequate for some applications (e.g., 
precision stellar photometry, cosmic shear measurement, etc.); in particular, the PSF width variation
is directly related to shear calibration (dilution correction) biases of cosmic shear measurements. 

\subsection{Comparison with TinyTim}
TinyTim is a software package for generating simulated PSFs for various instruments
installed on $HST$. The diffraction pattern is modeled by careful understanding of
the telescope's aperture and the optical path difference functions. Because the current
publicly available version of TinyTim is also capable of modeling
field-dependent variations in aberrations and charge diffusion for a full set of ACS filters
at different focus offsets, in principle it can obviate our empirical efforts to model
PSFs $if$ the results are consistent with stellar observations.

In order to compare TinyTim PSFs with those of real observations, we
generated PSF templates with TinyTim by varying focus values and star positions.
We changed the focus values at the $1 \mu m$ interval from $-10 \mu m$ to $+4 \mu m$, and
for a given focus we placed stars at the $\sim 125$ pixel interval, uniformly covering the WFC detector.
Because the final products by TinyTim are the ACS PSFs in a distorted frame, we
applied drizzle with the Lanczos3 kernel (the choice that we also made for the observation)
to simulate the geometric distortion correction effect.

We determined the ``focus'' of the observation in Figure~\ref{fig_starfield_image} to be
$-7 \mu m$ by searching for the TinyTim PSF template that best matched the observed 
ellipticity variation (shown in
the left panel of Figure~\ref{fig_ellipticity_recovery}). The first panel of Figure~\ref{fig_tiny_performance}
shows the ellipticity pattern for the observation predicted by TinyTim. Comparison with the
left panel of Figure~\ref{fig_ellipticity_recovery} gives a visual impression that TinyTim PSFs can reproduce
the global feature of the ACS PSF variation. However, when examined star-by-star, the TinyTim PSFs give large
systematic residuals (second panel); in general TinyTim stars appear to have more vertical elongation (i.e., smaller 
$\epsilon_{+}$). Obviously, these large systematic residuals translate into the high amplitudes of
ellipticity correlations (third panel). The PSF width variation (fourth panel)
predicted by TinyTim closely resembles the observed pattern
(see Figure~\ref{fig_psf_width} for comparison). We note, however, that 
the PSF widths of the TinyTim stars are systematically smaller ($\sim2$\%) than the observed
values.

The large systematic residuals in ellipticity are worrisome. This non-negligible
discrepancy between the TinyTim prediction and the observation suggests that
TinyTim ACS PSFs cannot be directly applied to ACS observations if PSF
anisotropy is of critical concern. For example, most weak-lensing studies draw
lensing signal from faint, small galaxies that are only slightly larger than
instrument PSFs. The large residual ellipticities shown in Figure~\ref{fig_tiny_performance}
can mimic (false) lensing signals.

\begin{figure}
\plotone{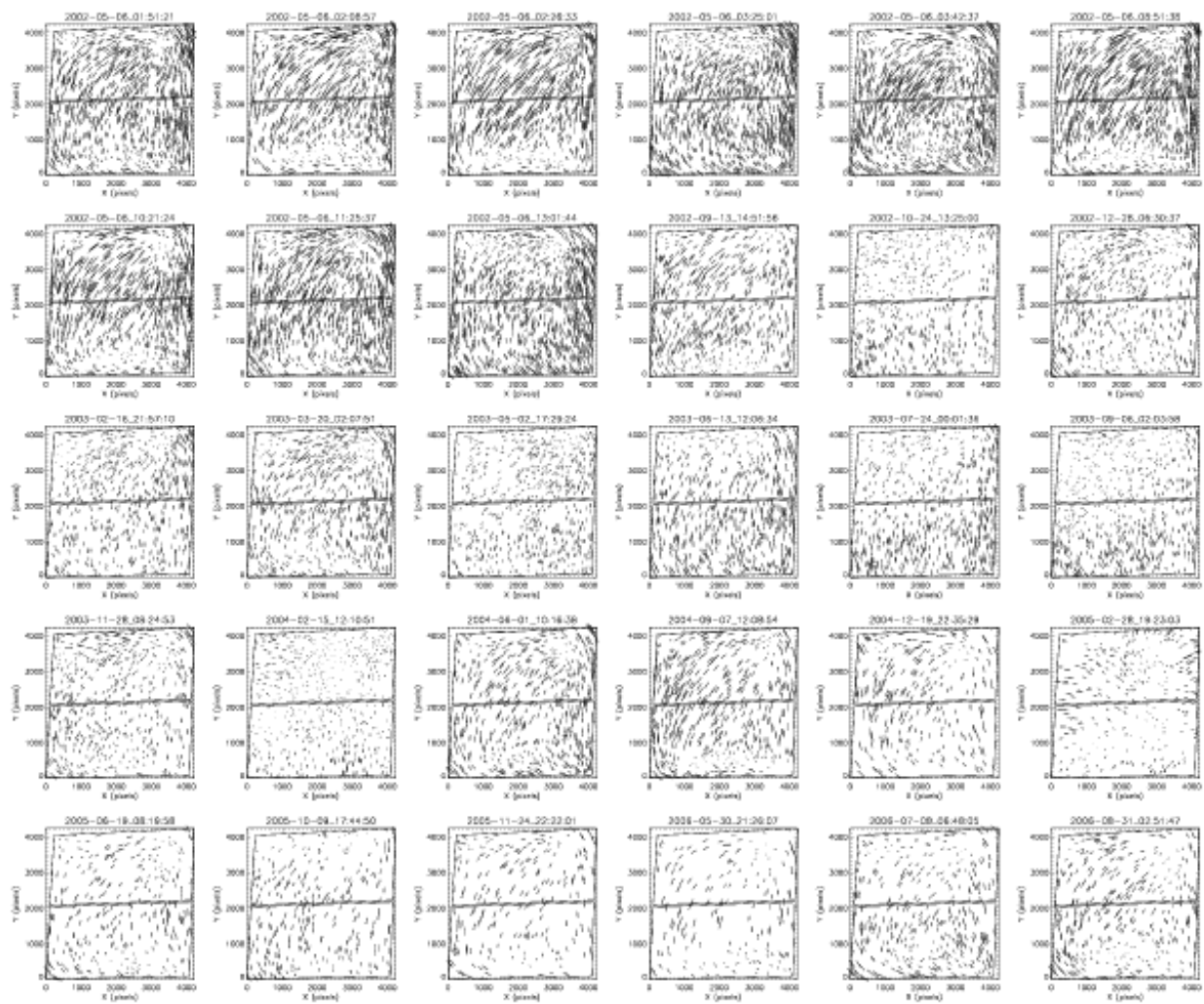}
\caption{Time- and position-dependent PSF ellipticity variation observed in 30 different
F435W exposures. Time increases to the right and to the bottom. The exposures are not homogeneously sampled in time
(The first 9 exposures are taken at different times on the same observation date). It is clear that the PSF ellipticity pattern
varies quite significantly in both direction and magnitude.
\label{fig_many_pattern}}
\end{figure}

\begin{figure}
\plotone{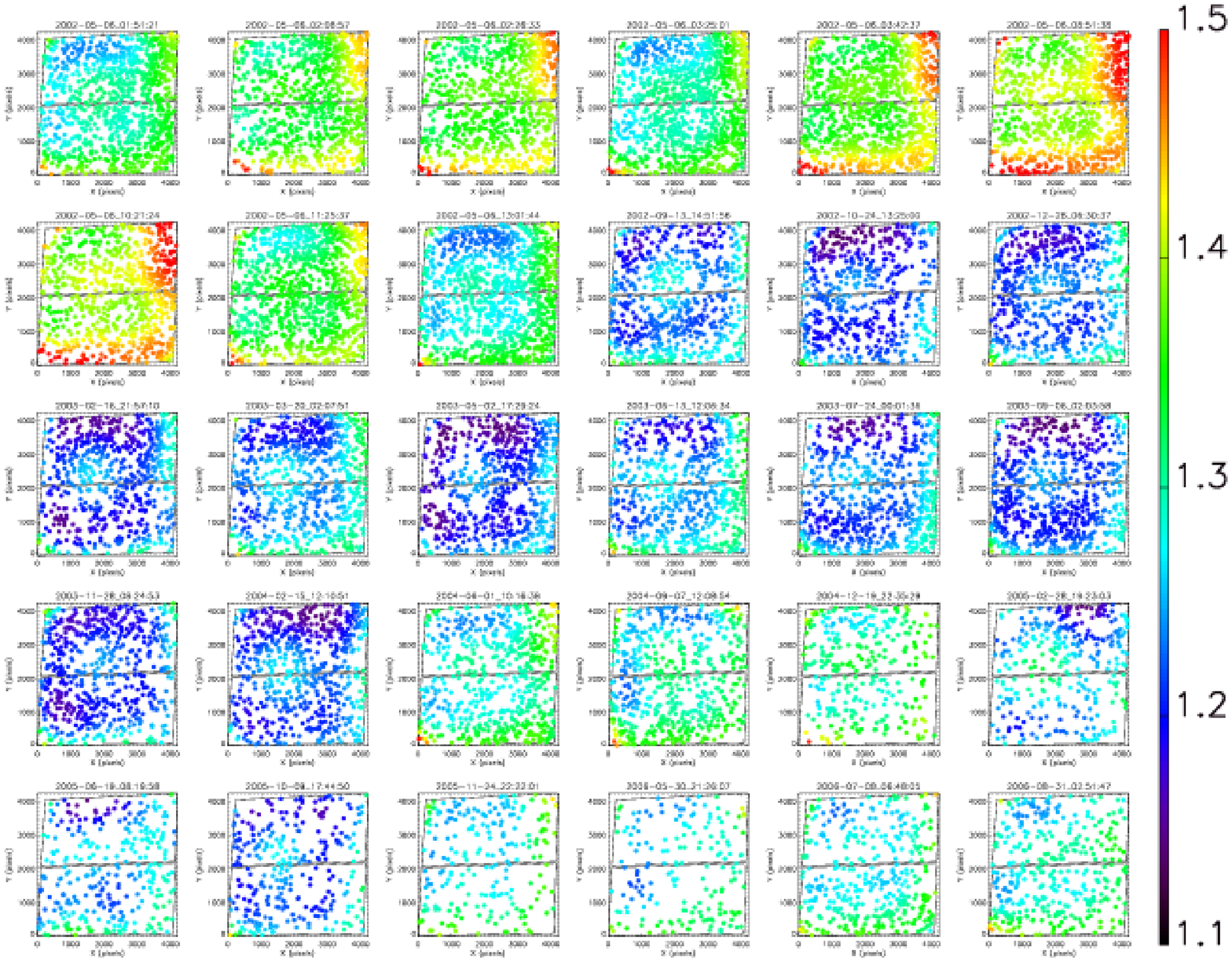}
\caption{Time- and position-dependent PSF width variation in the F435W exposures shown in Figure~\ref{fig_many_pattern}.
We arrange the frames in the same way as in Figure~\ref{fig_many_pattern}. We determine the widths of the PSFs with 
Equation~\ref{eqn_psf_width}. Comparison with Figure~\ref{fig_many_pattern} confirms that the average PSF widths per pointing are in general
proportional to the average magnitudes of ellipticities (i.e., size of whiskers). 
\label{fig_many_pattern_width}}
\end{figure}

\begin{figure}
\plotone{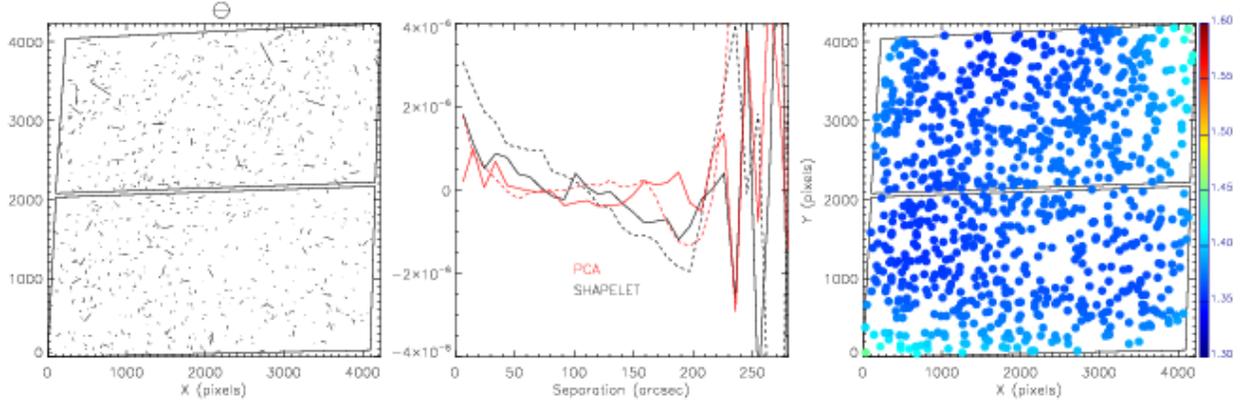}
\caption{Shapelet performance of ACS PSF representation. The residual ellipticity distribution in the left panel
shows the ellipticity residuals between the observed stars and the shapelet model. The middle panel displays
the spatial correlation of the residual ellipticity as a function of separation. Solid and dashed lines
are for $\xi_{+}$ and $\xi_{\times}$, respectively. We show the shapelet description of the PSF width variation
in the right panel. For the description of the comparison with the PCA results, see the text.
\label{fig_shapelet_performance}}
\end{figure}

\begin{figure}
\plotone{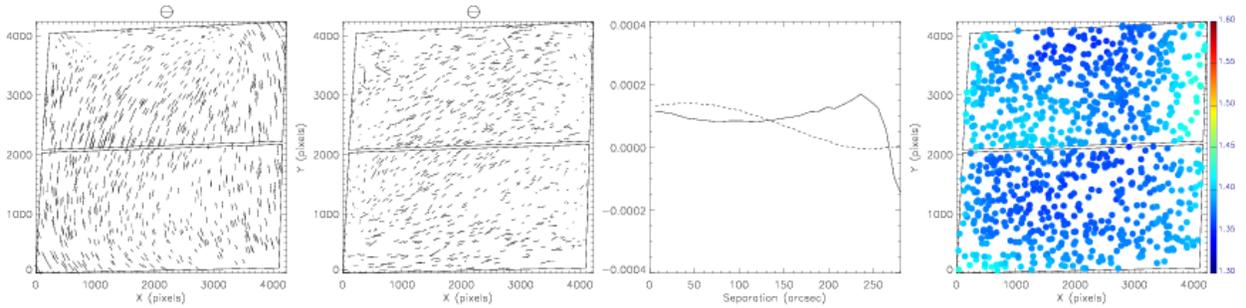}
\caption{TinyTim modeling of the PSF for the dataset J8C0D1051.
We find that the TinyTim PSFs generated with an input focus of $-7 \mu m$
best matches the observation. The ellipticity pattern of the stars predicted
by TinyTim (first panel) gives the visual impression that TinyTim PSFs can reproduce
the global pattern of the ACS PSF variation. However, when examined star-by-star
(second panel), the TinyTim PSFs give large systematic residuals (in general, 
TinyTim stars appear to have more vertical elongation). Obviously, these large 
systematic residuals translate into the high amplitudes of ellipticity
correlations (third panel), which are $\sim3$ orders of magnitude higher
than those from the PCA or shapelet approaches. The PSF width variation (fourth panel)
predicted by TinyTim closely resembles the observed pattern
(see Figure~\ref{fig_psf_width} for comparison). We note, however, that 
the PSF widths of the TinyTim stars are systematically smaller ($\sim2$\%) than the observed
values.
\label{fig_tiny_performance}}
\end{figure}

\begin{figure}
\plotone{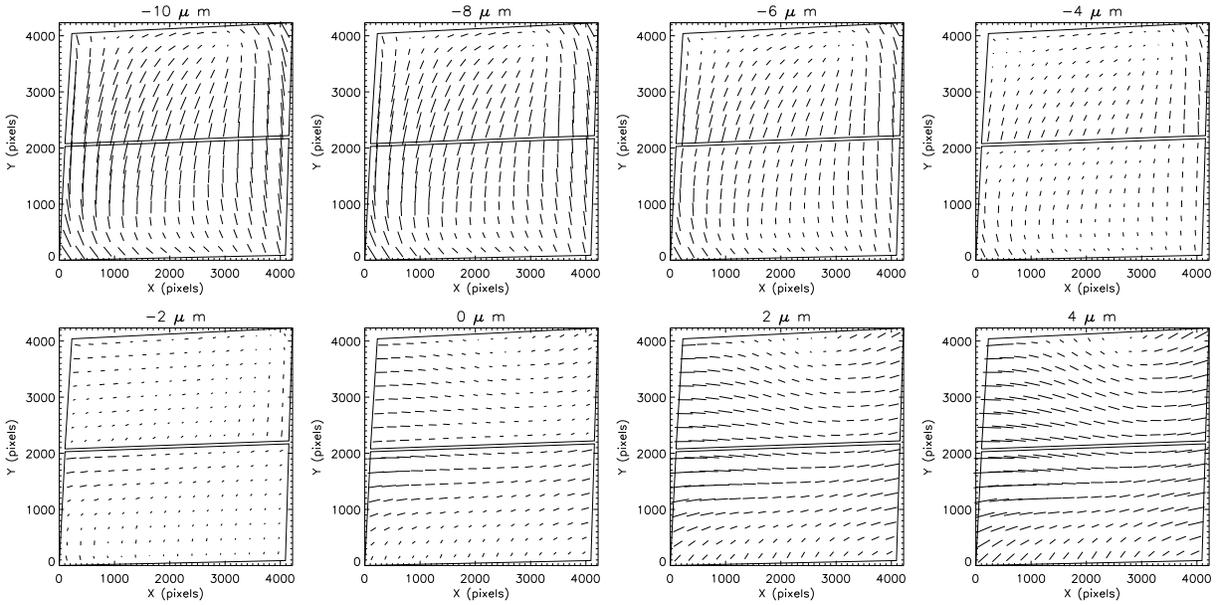}
\caption{PSF Ellipticity pattern predicted by TinyTim for different $HST$ focus values. We created
an array of $16\times16$ PSFs for each focus value and measure the ellipticity of these artificial
PSFs. The overall patterns somewhat resemble the ones in stellar observations (Figure~\ref{fig_many_pattern}).
However, on small scales there are important discrepancies between the TinyTim prediction and the observations. 
For example, the discontinuities across the two WFC chips do not seem to exist in observed PSFs.
\label{fig_many_tinytim}}
\end{figure}

\begin{figure}
\plotone{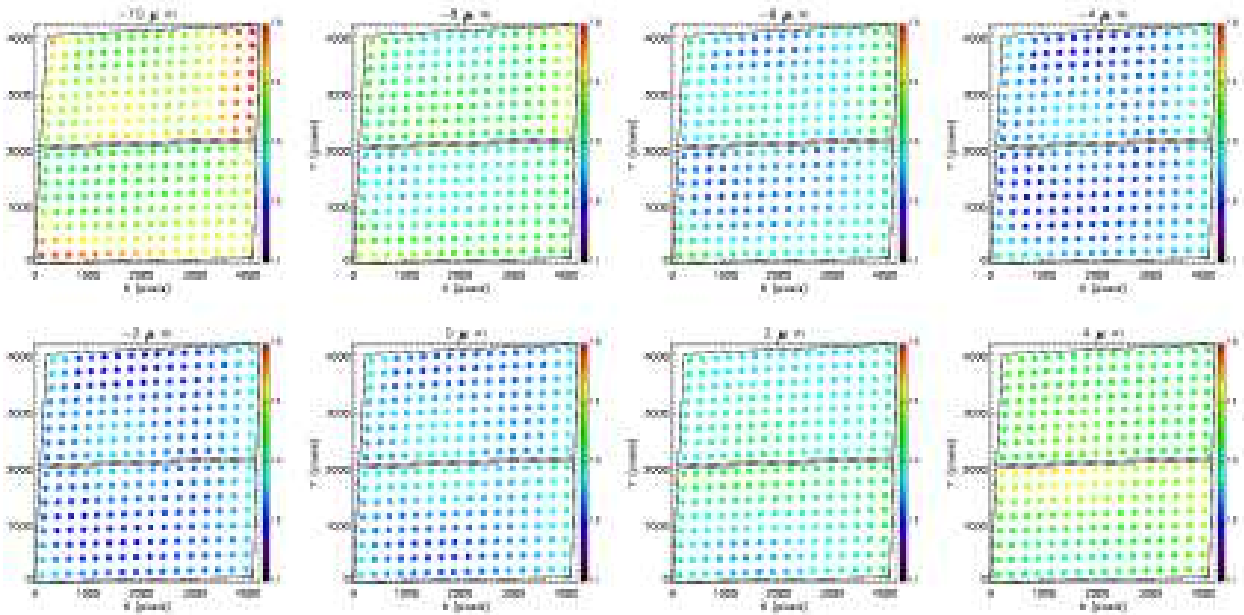}
\caption{Same as in Figure~\ref{fig_many_tinytim} except that here ACS/WFC PSF width variation patterns 
predicted by TinyTim are plotted instead. As in the case for the ellipticity comparison,
the overall patterns look similar to the ones in stellar observations (Figure~\ref{fig_many_pattern}).
However, non-negligible discrepancies between the TinyTim prediction and the observations exist on small
scales.
\label{fig_many_tinytim_width}}
\end{figure}

Of course, this TinyTim vs. observation mismatch is not confined to this
particular observation (J8C0D1051). To compare with the observed PSF patterns presented
in Figure~\ref{fig_many_pattern} and~\ref{fig_many_pattern_width} for F435W, we generate TinyTim PSF 
ellipticities of the same filter
for the focus values ranging from -10 $\mu$ m to $+4 \mu$ m in Figure~\ref{fig_many_tinytim} and 
~\ref{fig_many_tinytim_width}.
Again, we emphasize that
the TinyTim PSFs reproduce the global feature of the PSF variation; as described by Krist (2003),
we note that at negative focus values, the PSFs are on average elongated
from lower-left to upper-right and at positive focus values the average elongation rotates
by 90\degr. However, a scrutiny reveals that on small scales there exist some
important discrepancies similar to the ones already demonstrated in Figure~\ref{fig_tiny_performance}

One additional discrepancy deserving our attention is the feature near the gap between the two CCDs.
A conspicuous discontinuity in ellipticity is observed
in the TinyTim predictions (especially from the focus offset of $-6\mu$m to $-2\mu$m)
whereas in the observed PSFs the ellipticity change across the gap
appears to be continuous.  
The discontinuity appears because the TinyTim
assumes that there is a height offset of $0.02 \mu m$ between WFC1 and WFC2 based on
the previous focus-monitoring program results (Krist 2003).
Rhodes et al. (2007) also noticed this discrepancy between the observed stars in the COSMOS field
and the TinyTim model stars, and attributed the absence of this discontinuity across the chip
in observations to CTE degradations. This is a plausible explanation considering
that the CTE-induced charge trailing in Y-axis can cause an increase in PSF ellipticity
along the same direction; the regions near the chip gap are farthest from the readout
registers and thus are subject to greatest charge trailing. 
Because Rhodes et al. (2007) relied on TinyTim for the correction of PSF effects in the COSMOS field, 
they introduced some empirical multiplicative factors to their
quadrupole moment measurements in order to improve the agreement between TinyTim and COSMOS field stars.

However, although we observe that in certain situations the CTE degradation can lead to some smearing 
of PSFs in $y$ direction, we attribute the TinyTim and the observation mismatch largely
to the imperfection of TinyTim rather than to the imperfection of the CCDs (i.e., CTE degradation) based
on the following points. First, such discontinuities as predicted by TinyTim are
absent or negligibly small in real observations not only in the latest data, but also in the earliest
data soon collected after the installation of ACS on Hubble in March 2002.
The CTE degradation is mainly caused by exposure to high-energy charged particles in the space environment, and
thus the degradation grows with time.
If the effect of the CTE degradation is indeed the cause of the absence of
discontinuity in ellipticity pattern across the gap, we should have observed in early ACS data
the smallest discrepancies between TinyTim and real PSFs (i.e., the largest
discontinuity across the gap in observed PSFs).
Second, even if the charge trailing elongates PSFs, it cannot explain the absence of the discontinuity in
real observations. Because the CTE degradation is largest for the regions farthest from
the readout registers, the CTE-induced elongation should be equally greatest at the top
of WFC2 and at the bottom of WFC1. Therefore, the effect cannot reduce the discontinuity 
that TinyTim predicts between the two regions; if the CTE-induced elongation is unrealistically very large, 
it can give a false visual impression that the discontinuity is reduced.
Third, because the CTE degradation is supposed to elongate faint sources much more severely than bright ones, 
it is not probable
that the ellipticity patterns in Figure~\ref{fig_many_pattern} made from very high S/N stars
are severely affected by the charge trailing.
Finally, we don't expect to observe such substantial CTE charge trailing as to cause the large discrepancy between
the COSMOS stars and the TinyTim PSFs,  considering the relatively high sky background level ($>40$ $e^{-}$) 
of the COSMOS field. The background photons are supposed to fill charge traps and substantially mitigate 
the CTE degradation (Riess and Mack 2004). Therefore it is difficult to imagine that the CTE degradation 
selectively elongates
the bright COSMOS stars at the top of WFC2 and make the ellipticity change across the gap look 
artificially continuous.

\begin{figure}
\plotone{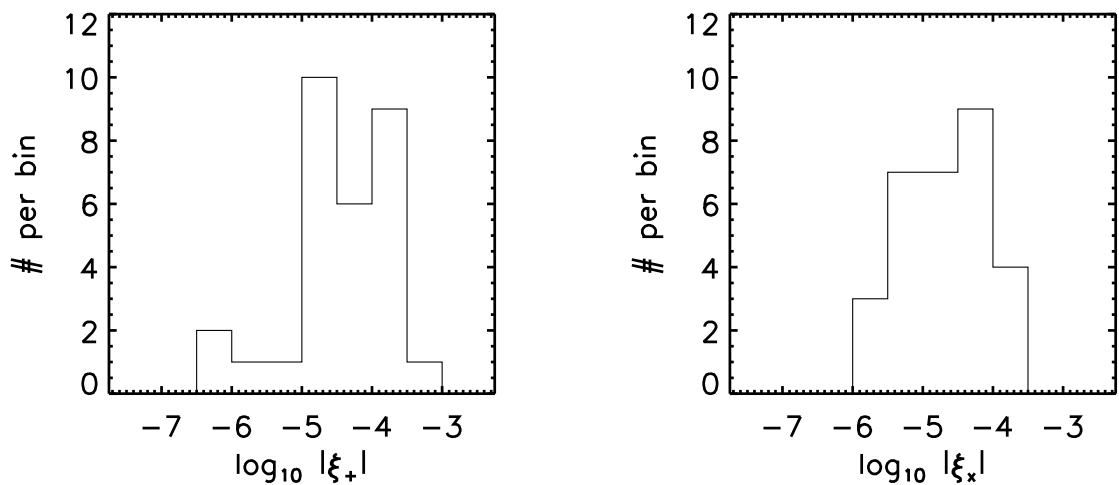}
\caption{PSF fitting with TinyTim. We display the distribution of the
mean residual ellipticity correlation for $\xi_{+}$ (left) and $\xi_{\times}$ (right)
after fitting TinyTim PSF to the exposures in Figure~\ref{fig_many_pattern}
\label{fig_tiny_hist}}
\end{figure}

\begin{figure}
\plotone{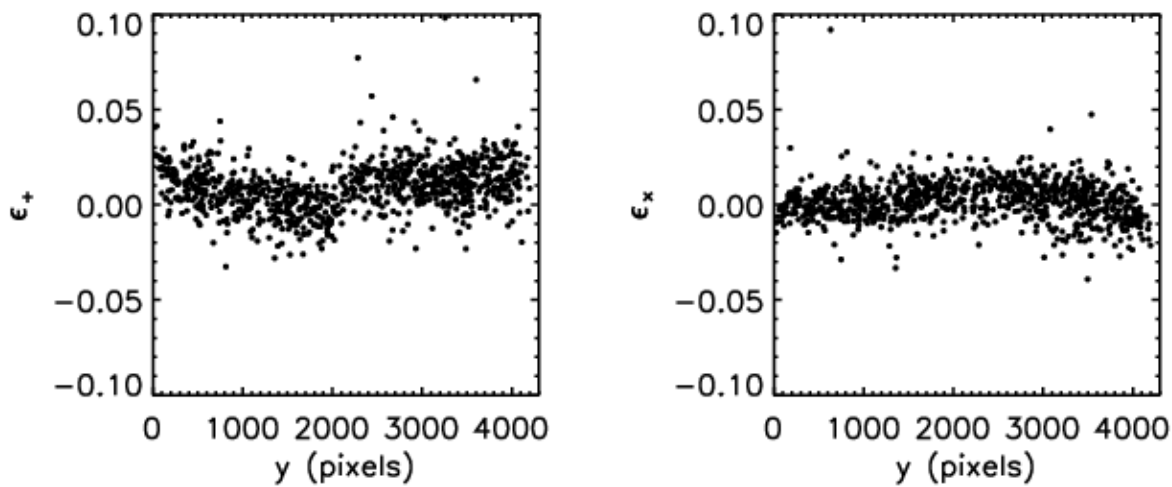}
\caption{TinyTim's ellipticity discontinuity across the gap between the two WFC chips.
The residual $\epsilon_{+}$ components (left) show a sudden, distinct
discontinuity of $\sim0.02$ at $y\sim2000$ whereas the feature is hard to identify in
the $\epsilon_{\times}$ residuals (right).
These residual ellipticities are evaluated after fittings TinyTim PSFs to
the first exposure shown in Figure~\ref{fig_many_pattern} 
(taken on 6 May 2002 at 1:51:21 UT).
\label{fig_tiny_discontinuity}}
\end{figure}

In Figure~\ref{fig_tiny_hist}, we display the distribution of the mean residual
ellipticity correlation for $\xi_{+}$ (left) and $\xi_{\times}$ (right)
after fitting TinyTim PSF to the 30 exposures in Figure~\ref{fig_many_pattern}.
We used the ellipticities of all the available high S/N stars in the fields ($200\sim900$)
to find the matching TinyTim PSF templates.
Not surprisingly, only a small fraction of the results give reasonably small
residual ellipticity correlation ($\lesssim 10^{-5}$). We observe that
none of the 30 exposures finds the TinyTim PSF template that
yields residual correlation of $\lesssim 10^{-6}$ for both $\xi_{+}$ and $\xi_{\times}$
simultaneously. Moreover, inspection of the $\epsilon_{+}$ residuals as a 
function of y-axis always shows a sudden, distinct discontinuity of $\sim0.02$
regardless of the observation epoch. We show one such example
in Figure~\ref{fig_tiny_discontinuity}, where we arbitrarily select the first exposure 
in Figure~\ref{fig_many_pattern} (taken on 6 May 2002 at 1:51:21 UT). 

In addition to the aforementioned discrepancies between TinyTim and observed PSFs, we also
point out here that the current version of TinyTim does not model a strong
scatter along the CCD serial readout direction at long wavelengths ($>8000$\AA). This horizontal
pattern (left panel of Figure~\ref{fig_psf_scatter}) is caused by an anti-halation layer introduced
between the CCD and its glass substrate (Sirianni et al. 1998), which is effective
at suppressing a near IR halo. The feature, which contains $\sim20$\% of the total PSF flux
at $1\mu m$, is strong in F850LP, and also visible in F814W whose transmission curve
truncates at 9300 \AA. The feature appears to enhance the existing horizontal diffraction spikes
particularly on the left-hand side of the core. However, unlike the real diffraction spikes, this scattering
feature penetrates deep into the PSF core, substantially increasing the PSF ellipticity
along $x$ direction (right panel of Figure~\ref{fig_psf_scatter}).

\subsection{How to Select the Right PSF Template}

Extracting the PSF information from stellar observations to construct
a library of PSF templates is one thing, but
finding a matching PSF template for a given science image is quite another.
Most of the existing stellar observations in the $HST$ archive are taken
in short exposures ($30\sim60$s) whereas typical science observations
require integration of one or more orbits. Particularly, weak-lensing
analysis of distant clusters ($z\gtrsim0.8$) needs multi-orbit
integration (with some dithering pattern) to achieve the aimed depth and
field of view.

Therefore, we must justify that the PSF template compiled from these short 
exposure observations can reliably represent the PSF pattern for the long-exposure 
science image, which should contain the intra-orbit focus variation.
Fortunately, previous studies (Jee et al. 2005a; Schrabback et al. 2007) support
the fact that the short-time exposure PSF pattern can
adequately serve as a mean PSF for long-exposure science observations.
Rhodes et al. (2007) also claim that the TinyTim
PSFs even at a fixed focus value can nicely represent the COSMOS stars
except for the aforementioned pattern in the detector center.

Having accepted that the short-exposure PSFs can properly serve as average PSFs
for long-exposure data, we can straightforwardly handle multi-orbit data
by finding a matching PSF template for each exposure. This method is proposed
by Jee et al. (2007), Schrabback et al. (2007), and Rhodes et al. (2007).
Jee et al. (2007) implemented the idea of finding a matching template for
each exposure from visual inspection of the star whiskers. This manual
procedure soon becomes prohibitively time-consuming as the number of exposures
increases. Rhodes et al. (2007) and Schrabback et al. (2007) suggested an automation
of the procedure by fitting the ellipticities of the PSF model (whether derived
from TinyTim or archival stellar fields) to the ellipticities of the stars in the target
field. We find that this method is a workable solution in general, but can be
improved by fitting the sizes of the PSFs, as well as the ellipticities. 
The merit is due to the observation that the sizes of the PSFs are also sensitive to the
focus of $HST$.
Therefore, our best fitting PSF template minimizes the following $\chi^2$:
\begin{equation}
\chi^2 = \sum \left [ \frac { (Q_{11}-Q'_{11})^2 }{ \sigma^2_{Q_{11}} } +
                      \frac { (Q_{22}-Q'_{22})^2 }{ \sigma^2_{Q_{22}} } +
                      \frac { (Q_{12}-Q'_{12})^2 }{ \sigma^2_{Q_{12}} } \right ] \label{eqn_psf_match}
\end{equation}
\noindent
where $Q_{ij}$ and $Q'_{ij}$ are the measurements of the stars in the science image and
the predicted values at the same locations in the model, respectively.
In the current study the uncertainties of the moments in Equation~\ref{eqn_psf_match} were evaluated 
from Monte Carlo simulations; alternatively, one can use analytic approximations (e.g., Goldberg \& Natarajan 2002).

\begin{figure}
\plotone{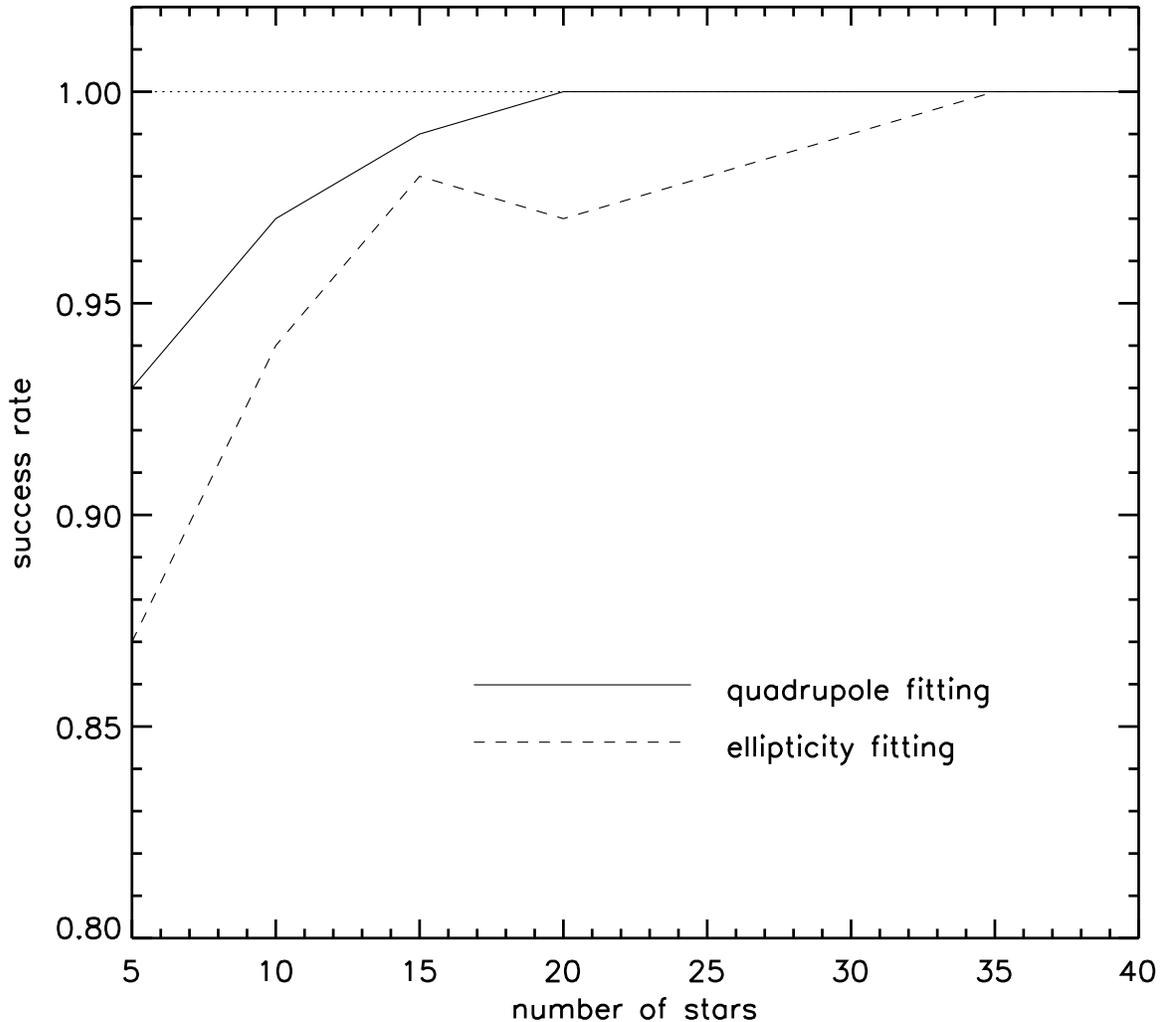}
\caption{Reliability test for PSF fitting as a function of the number of stars used. We
randomly selected a small number of stars (out of $\sim800$) from the dataset J8C0D1051 and found a
matching template from our library based on their ellipticities or alternatively
quadrupole moments. We iterated 100 times for a given number of stars and examined how
many incidences fall to the category of ``success''. We consider the incidence as failure 
if the absolute value of the resulting ellipticity correlation becomes greater than $10^{-5}$, which
is a very conservative choice. The number of failure incidences for quadrupole fitting
is significantly lower than for ellipticity fitting.
\label{fig_psf_fitting_error}}
\end{figure}

\begin{figure}
\plottwo{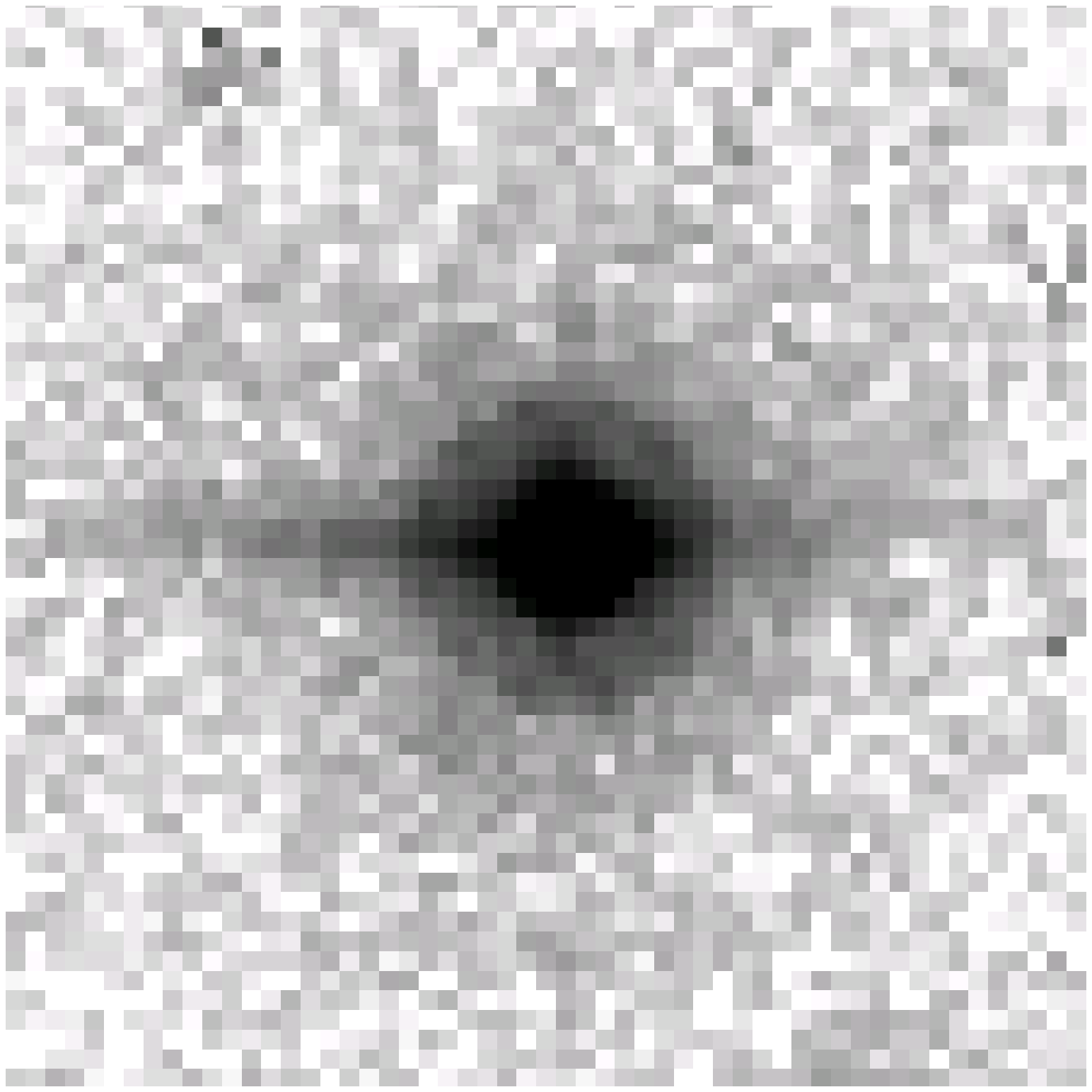}{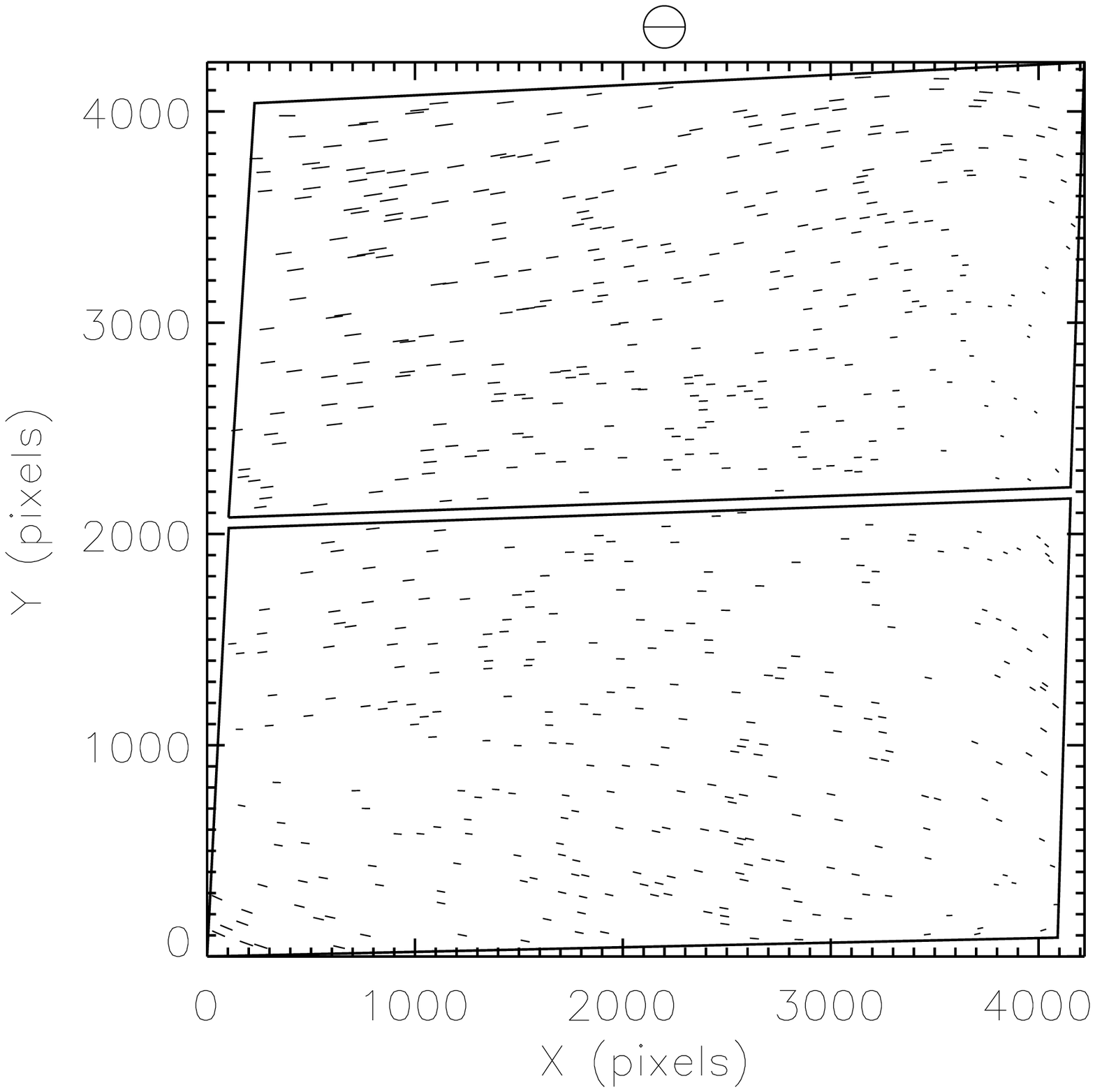}
\caption{(a) Red light scatter in F850LP due to the anti-halation layer. The metal coating that was applied to the front-side
of the WFC CCD for the suppression of near IR halos creates a horizontal scattering feature
for long wavelength photons ($>8000$\AA). The feature enhances the existing horizontal diffraction spikes
particularly on the left-hand side of the core. (b) A typical PSF ellipticity pattern in F850LP. This horizontal
scattering dominantly affects the ellipticity pattern in F850LP.
\label{fig_psf_scatter}}
\end{figure}

We examined the reliability of the above PSF fitting by randomly drawing a small number of
stars from the catalog of J8C0D1051 and fitting the template PSF from our library to
these stars. The catalog of J8C0D1051 contains $\sim870$ stars and as already shown in 
Figure~\ref{fig_ellipticity_recovery} a small fraction ($\sim6$\%) of
these stars ($\sim150$) are noisy (residual ellipticity greater than 0.02). 
We did not discard these noisy stars in the random selection because we want to
simulate realistic cases where it is hard to judge which stars are noisy.
The reliability of the PSF fitting of course depends on the properties and the locations 
of the selected stars. Therefore, we iterated 100 times for a given number of stars to
even out the selection effect. We evaluated the quality of the fitting using the resulting
ellipticity correlation functions. We consider the fitting as failure if the absolute
value of the mean correlation is greater than $10^{-5}$, which is a very conservative
choice (cosmic shear signals are of the order $10^{-4}$. however, they are
measured from galaxies whose ellipticity correlations due to the PSF correlations
are somewhat diluted). Figure~\ref{fig_psf_fitting_error} displays the simulation result.
A few points are worthy to be discussed. First, the success rate is still high even 
when only 5 stars are used (7 and 13 out of 100 failures
for quadrupole and ellipticity fitting, respectively). Second, the number of failure 
incidences for quadrupole fitting is significantly lower than for ellipticity fitting
(approximately a factor of two less incidences for fewer than 15 stars).
Finally, we note that even for the ``failure'' incidences the resulting ellipticity
correlation is only moderately high ($\sim10^{-4}$ or less).

\section{CONCLUSIONS \label{section_conclusion}}
We showed that the time- and position-dependent ACS/WFC PSF can be robustly described through PCA. The PCA
technique allows us to perform orthogonal expansion of the observed PSFs with as few as 20 eigen-PSFs
derived from the data themselves. This method is superior to our previous shapelet-based decomposition
of the PSFs, capturing more details of the diffraction pattern of the instrument PSF.
By interpolating the position-dependent variation of the eigen-PSFs with 5th order polynomials, we are able
to recover the observed pattern of the PSF ellipticity and width variation. Although the TinyTim software provides
a good approximation of the observed PSFs, we demonstrate that there are some important mismatches between
the TinyTim prediction and the real PSFs, which cannot be attributed to CTE degradation of WFC over time.
The CTE charge trailing effect should be negligible for these bright high S/N stars, and we do not
observe any long-term variation of the pattern (i.e., increasing elongation in parallel read-out direction with time) 
due to the CTE degradation. 
Because typical science observations require integration of one or more orbits in broadband filters, 
the background levels are high ($\sim200$ $e^{-}$ for integration of one orbit).
These high background photons are supposed to fill the charge traps and thus mitigate the CTE effects.
Therefore, we argue that the CTE-induced elongation is not likely to limit the application of our
PSF models extracted from short-exposure observations to long-exposure science images.

We have compiled WFC PSFs from $>400$ stellar field observations, which span a wide range of
$HST$ focus values. Although the current paper mainly deals with the ACS/WFC PSF issue in the
context of weak-lensing analysis, we believe that our PSF model can be used in a wide range of the astronomical data analyses 
where the knowledge of the position-dependent WFC PSF is needed (e.g., crowded field stellar photometry, 
robust profile fitting of small objects, weak-lensing analyses, etc.).

ACS was developed under NASA contract NAS5-32865, and this research was supported
by NASA grant NAG5-7697.

\begin{thebibliography}{}
\bibitem[Anderson \& King (2006)]{ak06} Anderson, J. \& King \ 2006, $Instrument$ $Science$ $Report$ $ACS$ 2006-01, Space Telescope Science
Institute
\bibitem[Bernstein \& Jarvis(2002)]{2002AJ....123..583B} Bernstein, G.~M., 
\& Jarvis, M.\ 2002, \aj, 123, 583 
\bibitem[Bromley et al.(1998)]{1998ApJ...505...25B} Bromley, B.~C., Press, 
W.~H., Lin, H., \& Kirshner, R.~P.\ 1998, \apj, 505, 25 
\bibitem[Connolly et al.(1995)]{1995AJ....110.1071C} Connolly, A.~J., 
Szalay, A.~S., Bershady, M.~A., Kinney, A.~L., \& Calzetti, D.\ 1995, \aj, 
110, 1071 
\bibitem[Goldberg \& Natarajan(2002)]{2002ApJ...564...65G} Goldberg, D.~M., 
\& Natarajan, P.\ 2002, \apj, 564, 65 
\bibitem[Hack et al.(2003)]{2003ASPC..295..453H} Hack, W., Busko, I., \&
Jedrzejewski, R.\ 2003, ASP Conf.~Ser.~295: Astronomical Data Analysis
Software and Systems XII, 295, 453
\bibitem[Hershey, J. (1997)]{Hershey1997} Hershey, J.\ 1997, Modeling HST Focal-Length Variations V.1.1, SESD-97-01
\bibitem[Heymans et al.(2005)]{2005MNRAS.361..160H} Heymans, C., et al.\ 
2005, \mnras, 361, 160 
\bibitem[Jarvis \& Jain(2004)]{2004astro.ph.12234J} Jarvis, M., \& Jain, 
B.\ 2004, ArXiv Astrophysics e-prints, arXiv:astro-ph/0412234 
\bibitem[Jee et al.(2005)]{2005ApJ...618...46J} Jee, M.~J., White, R.~L.,
Ben{\'{\i}}tez, N., Ford, H.~C., Blakeslee, J.~P., Rosati, P., Demarco, R.,
\& Illingworth, G.~D.\ 2005a, \apj, 618, 46
\bibitem[Jee et al.(2005)]{2005ApJ...634..813J} Jee, M.~J., White, R.~L., 
Ford, H.~C., Blakeslee, J.~P., Illingworth, G.~D., Coe, D.~A., \& Tran, 
K.-V.~H.\ 2005b, \apj, 634, 813 
\bibitem[Jee et al.(2006)]{2006ApJ...642..720J} Jee, M.~J., White, R.~L., 
Ford, H.~C., Illingworth, G.~D., Blakeslee, J.~P., Holden, B., \& Mei, S.\ 
2006, \apj, 642, 720 
\bibitem[Jee et al.(2007)]{2007ApJ} Jee, M.~J., 
Ford, H.~C., Illingworth, G.~D., , White, R.~L., Broadhurst, T.~J., Coe, D.~A.,
Meurer, G.~R., van der Wel, A., Benitez, N., Blakeslee, J.~P., Bouwens, R.~J.,
Bradley, L., Demarco, R., Homeier, N. ~L. , Martel, A.~R., \&  \& Mei, S.\ 
2007, \apj, 661, 728
\bibitem[Krist(2003)]{krist03} Krist, J.\ 2003, $Instrument$ $Science$ $Report$ $ACS$ 2003-06, Space Telescope Science
Institute
\bibitem[Lauer (2002)]{2002astro.ph} Lauer, T.~R. \ 
2002, ArXiv Astrophysics e-prints, arXiv:astro-ph/0208247
\bibitem[Lupton et al.(2001)]{2001ASPC..238..269L} Lupton, R., Gunn, J.~E., 
Ivezi{\'c}, Z., Knapp, G.~R., \& Kent, S.\ 2001, Astronomical Data Analysis 
Software and Systems X, 238, 269 
\bibitem[Madgwick et al.(2003)]{2003ApJ...599..997M} Madgwick, D.~S., et 
al.\ 2003, \apj, 599, 997 
\bibitem[Press et al. (1992)]{1992press} Press, W.H., Teukolsky, S.A., Vetterling, W.T., Flannery, B.P. 1992,
Numerical Recipes. Cambridge (Cambridge University Press)
\bibitem[Refregier(2003)]{refregier03} Refregier, A.\ 2003, \mnras, 338, 35
\bibitem[Rhodes et al.(2007)]{2007astro.ph..2140R} Rhodes, J.~D., et al.\ 
2007, ArXiv Astrophysics e-prints, arXiv:astro-ph/0702140 
\bibitem[Riess and Mack (2004)]{rm04} Riess, A. \& Mack, J. 2004, $Instrument$ $Science$ $Report$ $ACS$ 2004-006, Space Telescope Science
\bibitem[Schrabback et al.(2007)]{2007A&A...468..823S} Schrabback, T., et 
al.\ 2007, \aap, 468, 823  
\bibitem[Sirianni et al.(1998)]{1998SPIE.3355..608S} Sirianni, M., et al.\ 
1998, \procspie, 3355, 608 
\end{thebibliography}
\end{document}